\documentclass[aps,nofootinbib,showpacs,preprintnumbers,amsmath,amssymb]{revtex4}
\def\bes{\begin{subequations}}
\def\ees{\end{subequations}}
\def\be{\begin{equation}}
\def\ee{\end{equation}}
\def\bea{\begin{eqnarray}}
\def\eea{\end{eqnarray}}
\def\ba{\begin{eqnarray}}
\def\ea{\end{eqnarray}}
\def\bear{\begin{array}}
\def\eear{\end{array}}

\newcommand{\K}{{\widetilde {\cal K}}}
\newcommand{\bG}{{\overline{\Gamma}}}
\newcommand{\Br}{{\overline {\rm Br}}}

\usepackage{graphics}
\usepackage{graphicx}
\usepackage{dcolumn}
\usepackage{bm}
\usepackage{epsfig}
\usepackage{graphicx}
\usepackage{multirow}
\usepackage{dcolumn}
\usepackage{graphicx,epsfig}%

\begin{document}
\preprint{USM-TH-316}

\title{CP violations in $\pi^{\pm}$ Meson Decay}
\author{Gorazd Cveti\v{c}$^1$}
\author{C.~S.~Kim$^2$}
\email{cskim@yonsei.ac.kr}
\author{Jilberto Zamora-Sa\'a$^1$}
\affiliation{$^1$Department of Physics, Universidad T\'ecnica Federico Santa Mar\'ia, Valpara\'iso, Chile\\
$^2$Department of Physics and IPAP, Yonsei University, Seoul 120-749, Korea}

\date{\today}

\begin{abstract}
We study the pion decays with intermediate on-shell neutrinos $N$
into two electrons and a muon, 
$\pi^{\pm} \to e^{\pm} N \to e^{\pm} e^{\pm} \mu^{\mp} \nu$.
We investigate the branching ratios 
${\rm Br}_{\pm} = [\Gamma(\pi^- \to e^- e^- \mu^+ \nu) \pm \Gamma(\pi^+ \to e^+e^+\mu^-\nu)]/\Gamma(\pi^- \to {\rm all})$ and the
CP asymmetry ratio ${\cal A}_{\rm CP} = {\rm Br}_{-}/{\rm Br}_{+}$
for such decays, in the scenario with two different on-shell neutrinos.
If $N$ is Dirac, only the lepton number conserving (LC) decays contribute
(LC: $\nu= {\nu}_e$ or ${\bar \nu}_e$); if $N$ is Majorana, both LC and
lepton number violating (LV) decays contribute (LV: $\nu={\bar \nu}_{\mu}$
or $\nu={\nu}_{\mu}$). The results show that the CP asymmetry 
${\cal A}_{\rm CP}$ is in general very small, 
but increases and becomes $\sim 1$
when the masses of the two intermediate neutrinos get closer to each other,
i.e., when their mass difference becomes comparable with their decay width,
$\Delta M_N \not\gg  \Gamma_N$.
The observation of CP violation
in pion decays would be consistent with the existence of the well-motivated
$\nu$MSM model with two almost degenerate heavy neutrinos.
\end{abstract}


\maketitle

\section{Introduction}
\label{intr}

One of the outstanding issues in neutrino physics today is to
clarify the  Dirac or Majorana character of neutrino masses.
If neutrinos are Dirac particles, they must have right-handed
electroweak singlet components in addition to the known left-handed
modes; in such case lepton number remains as a conserved quantity.
Alternatively, if they are Majorana particles, they are indistinguishable
from their antiparticles, and the lepton number in the reactions involving them
may be violated. The nature of the neutrinos can be
discerned via detection of neutrinoless double beta decays ($0 \nu \beta \beta$)
in nuclei \cite{0nubb}, by considering specific scattering processes
\cite{scatt}, or by studying rare meson decays \cite{rmeson,JHEP}.
The experimental results to date are unable to distinguish between
these two alternatives.

Among the principal tasks in neutrino physics are the ascertainment of
the nature of the neutrino mass (Dirac or Majorana) and the
CP violation in the neutrino sector.
The measurement of neutrino oscillations \cite{Pontecorvo,oscatm,oscsol,oscnuc}
suggests that the first three
neutrinos are not massless but very light particles, with masses less than 1 eV.
If these light masses are produced via a seesaw \cite{seesaw} or related mechanism,
then the existence of significantly heavier neutrinos is expected.
Furthermore, there is a possibility of CP violation in the neutrino sector,
both if neutrinos are Dirac or Majorana particles. In the Majorana case,
though, the number of possible CP-violating phases in the PMNS matrix is larger.
If $n$ is the number of neutrino generations, the number of CP-violating phases
is $(n-1)(n-2)/2$ in the Dirac case, and $n(n-1)/2$ in the Majorana case,
cf.~Ref.~\cite{Bilenky}.

In this work, we investigate the possibility of measuring the CP asymmetry in the rare pion decays:
The CP violation in the neutrino sector can be measured by neutrino oscillations
\cite{oscCP}. However, here we consider a scenario in which CP violation
of the neutrino sector can be measured by investigating rare meson decays.
We consider a scenario of two additional, sterile,  almost degenerate neutrinos $N_j$ ($j=1,2$)
with masses $M_N \sim 10^2$ MeV. Such neutrinos are not typically predicted by
seesaw scenarios; nonetheless, there are models which predict such neutrinos
\cite{nuMSM,Shapo}, and they are not ruled out by experiments \cite{PDG2012,Atre}.

We note that the model, $\nu$MSM~\cite{nuMSM,Shapo},
proposes two almost degenerate Majorana neutrinos with mass
between $100$ MeV and a few GeV, in addition to a light Majorana 
neutrino of mass $\sim 10^1$ keV.
The existence of such neutrinos is strongly motivated, because
it can explain simultaneously the baryon asymmetry of the Universe,
the pattern of light neutrino masses and oscillations, and can provide 
a dark matter candidate -- cf.~\cite{nuMSMrev} for a review, 
and \cite{CDSh} for the allowed range of 
the sterile neutrinos in $\nu$MSM.\footnote{
The tentative evidence of a dark matter line, recently discussed in \cite{DM}, 
is well within the regime predicted in $\nu$MSM in  \cite{CDSh}. 
We thank Marco Drewes for bringing this point to our attention.}
The requirement that the lightest sterile neutrino be
the dark matter candidate reduces the parameters of the model
in such a way as to make the two heavier neutrinos nearly degenerate
in mass.

Recently CERN-SPS has proposed a search of such
heavy neutrinos, Ref.~\cite{CERN-SPS}, 
in the decays of $D$, $D_s$ mesons. 
We are interested in the question whether in such models
the CP violation in rare pion decays can be appreciable to cover 
the parameter space favored by theoretical models.

We investigate the rare decays of
charged pions into three charged leptons and a light neutrino,
with the two intermediate neutrinos $N_j$ in the decay being on-shell,
and we look for a possibility of detection of CP asymmetries in such decays.
The relevant processes are the
lepton number conserving (LC) processes
$\pi^{\pm} \to e^{\pm} N_j \to e^{\pm} e^{\pm} \mu^{\mp} \nu$ where
$\nu=\nu_e$ for $\pi^+$ and $\nu={\bar \nu}_e$ for $\pi^-$;
and the lepton number violating (LV) processes, where
$\nu={\bar \nu}_{\mu}$ for $\pi^+$ and $\nu={\nu}_{\mu}$ for $\pi^-$.
If the $N_j$ neutrinos are Dirac, only LC decays contribute.
If they are Majorana, both LC and LV decays contribute.
In our previous work \cite{JHEP}, we demonstrated that the
decay branching ratios for these processes are very small but
can be appreciable and could be measured in the future $\pi$ factories
where huge numbers of pions will be produced, if the heavy-light neutrino
mixing parameters are sufficiently large but still below the present
upper bounds. Moreover, we showed that the consideration of the
muon spectrum of these decays may allow us to distinguish whether the
intermediate neutrinos are Dirac or Majorana.

We will investigate the branching ratios
${\rm Br}_{\pm} \equiv [\Gamma(\pi^- \to e^- e^- \mu^+ \nu) \pm \Gamma(\pi^+ \to e^+e^+\mu^-\nu)]/\Gamma(\pi^- \to {\rm all})$ and the CP asymmetry
ratio ${\cal A}_{\rm CP} \equiv {\rm Br}_{-}/{\rm Br}_{+}$
of the mentioned rare processes
in the scenario of two intermediate on-shell neutrinos.
We demonstrate that there exist scenarios where this CP asymmetry can be
detected. In Sec.~\ref{sec:form} we outline the formalism for the calculation
of the various decay widths and branching ratios.
The details of the calculation are given in Appendix \ref{app1}.
In Sec.~\ref{sec:ACPsum} we derive the expressions for the
branching ratios
${\rm Br}_{\pm}$ and for the CP asymmetry ratio ${\cal A}_{\rm CP}$, and
present the numerical results.
Additional details are given in Appendix \ref{app2}.
In Sec.~\ref{sec:concl} we present the conclusions.

\section{The processes and formalism for the rare pion decays}
\label{sec:form}

\begin{figure}[htb] 
\begin{minipage}[b]{.49\linewidth}
\centering\includegraphics[width=65mm]{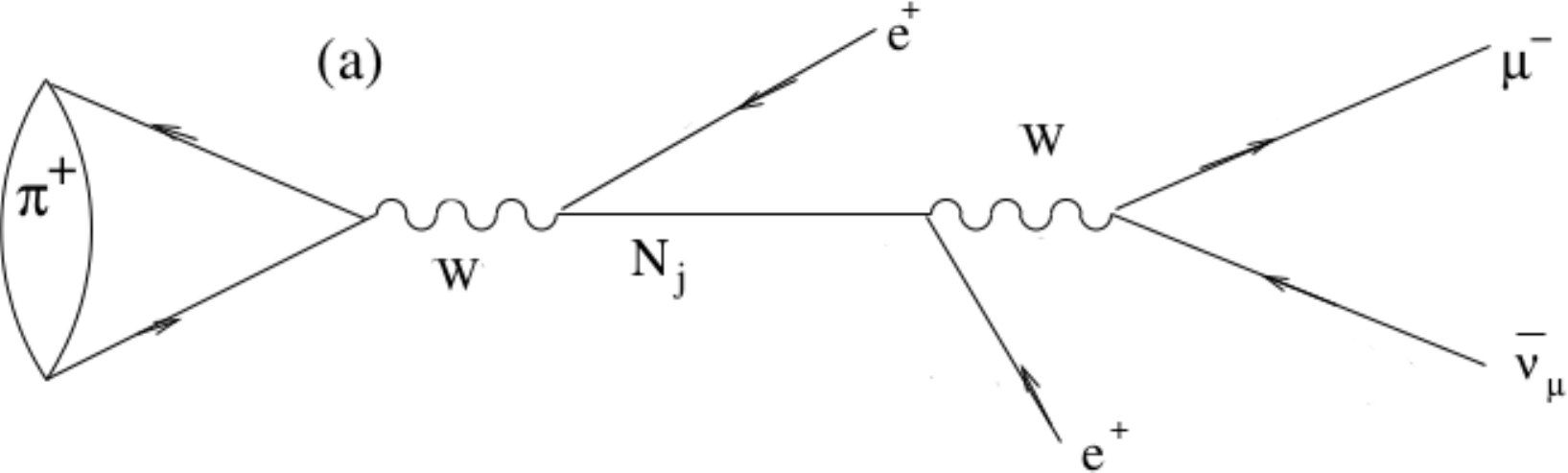}
\end{minipage}
\begin{minipage}[b]{.49\linewidth}
\centering\includegraphics[width=65mm]{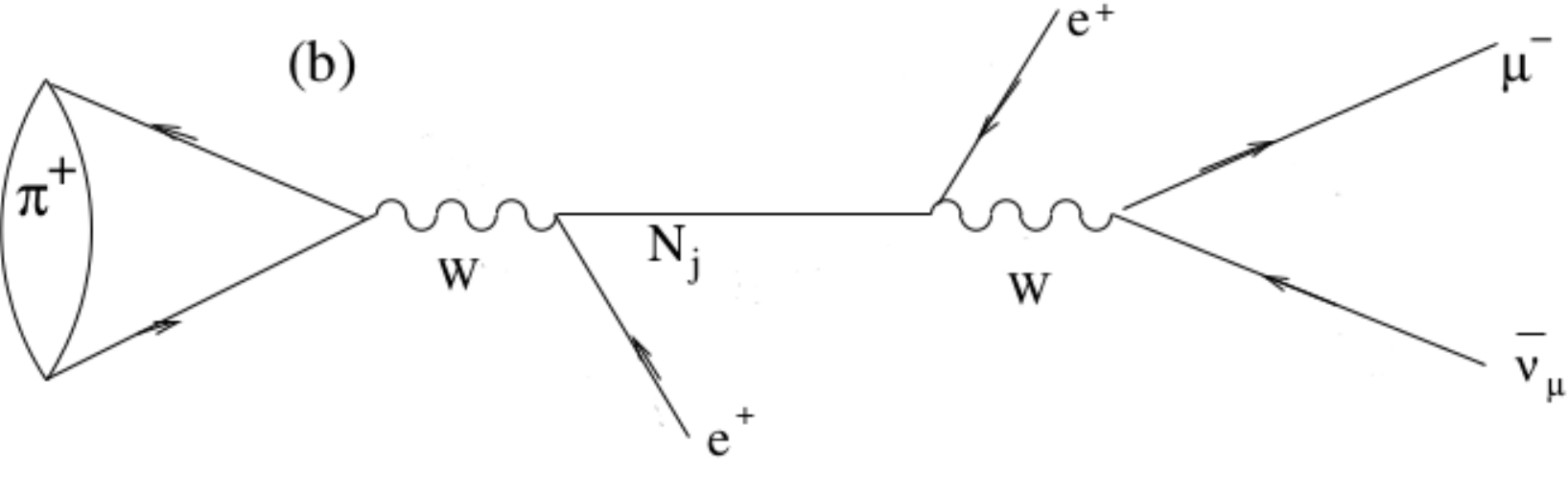}
\end{minipage}
\vspace{-0.4cm}
\caption{The lepton number violating (LV) process: (a) the direct (D) channel; (b) the crossed (C) channel.}
\label{FigLV}
 \end{figure}

\begin{figure}[htb] 
\begin{minipage}[b]{.49\linewidth}
\centering\includegraphics[width=65mm]{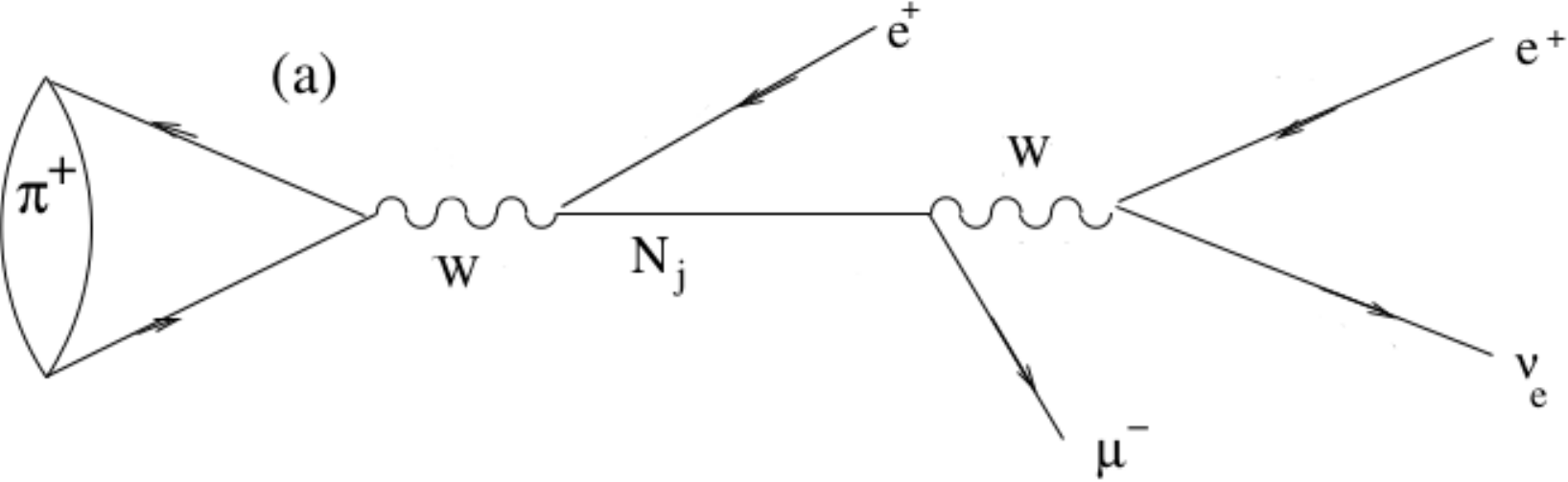}
\end{minipage}
\begin{minipage}[b]{.49\linewidth}
\centering\includegraphics[width=65mm]{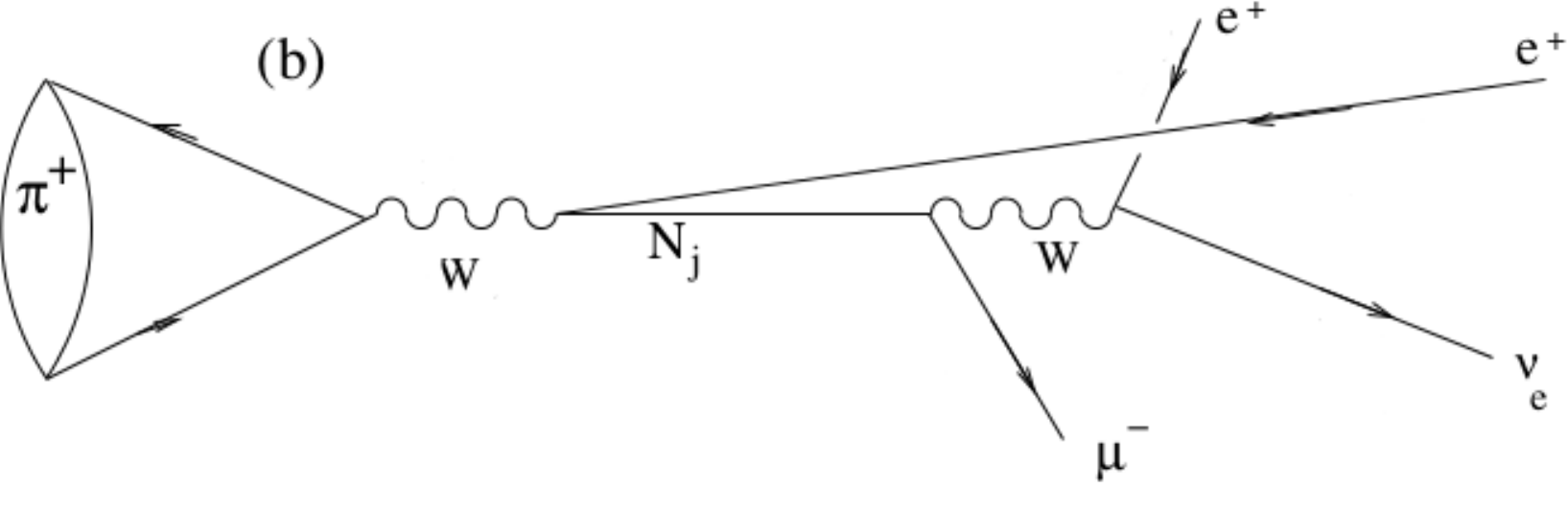}
\end{minipage}
\vspace{-0.4cm}
\caption{The lepton number conserving (LC) process: (a) the direct (D) channel; (b) the crossed (C) channel.}
\label{FigLC}
 \end{figure}

We consider the lepton number violating (LV) process, Fig.~\ref{FigLV}, and the
lepton number conserving (LC) process, Fig.~\ref{FigLC}.
We note that if the intermediate neutrinos $N_j$ ($j=1,2$) are Majorana, both processes
(LV and LC) take place; and if $N_j$ are Dirac, only the LC process takes place.

We will denote the mixing coefficient between the standard flavor neutrino $\nu_{\ell}$
($\ell = e, \mu, \tau$) and the heavy mass eigenstate $N_j$ as $B_{\ell N_j}$ ($j=1,2)$, i.e., this mixing element appears in the relation
\be
\nu_{\ell} = \sum_{k=1}^3 B_{\ell \nu_k} \nu_k +
\left( B_{\ell N_1} N_1 + B_{\ell N_2} N_2 \right) \ ,
\label{mix}
\ee
where $\nu_k$ ($k=1,2,3$) are the light mass eigenstates.
We adopt the phase conventions of the book Ref.~\cite{Bilenky}, i.e.,
all the CP-violating phases are incorporated in the PMNS matrix of mixing
elements.
The decay widths and asymmetries of these processes may become appreciable only if the
two intermediate neutrinos $N_j$ are on-shell, i.e., if
\be
(M_{\mu} + M_e) < M_{N_j} < (M_{\pi}-M_e) \ ,
\label{MNjint}
\ee
i.e., when the masses $M_{N_j}$ are within the interval (106.2 MeV, 139 MeV).

From now on, unless otherwise stated, we will use the simplified notations
for the decay widths of these rare processes:
\be
\Gamma^{(X)}(\pi^{\pm}) \equiv
\Gamma^{(X)}(\pi^{\pm} \to e^{\pm} e^{\pm} \mu^{\mp} \nu) \ ,
\quad (X={\rm LV, \ LC}) \ .
\label{not0}
\ee
The decay widths $\Gamma^{(X)}(\pi^{\pm})$ can be written in the form
\be
\Gamma^{(X)}(\pi^{\pm}) =  \frac{1}{2!} \frac{1}{2 M_{\pi}} \frac{1}{(2 \pi)^8}
\int d_4 \; | {\cal T}^{(X)}(\pi^{\pm}) | ^2 \ ,
\label{GX1}
\ee
where $1/2!$ is the symmetry factor due to two final state electrons,
and $d_4$ denotes the integration over the 4-particle final phase space
\be
d_4 =\left( \prod_{j=1}^2  \frac{d^3 {\vec p}_j}{2 E_e({\vec p}_j)} \right)
 \frac{d^3 {\vec p}_{\mu}}{2 E_{\mu}({\vec p}_{\mu})}
 \frac{d^3 {\vec p}_{\nu}}{2 |{\vec p}_{\nu}|}
\delta^{(4)} \left( p_{\pi} - p_1 - p_2 - p_{\mu} - p_{\nu} \right) \ ,
\label{d4}
\ee
and we denoted by $p_1$ and $p_2$ the momenta of $e^+$
from the left and the right vertex of the direct channels,
respectively (and for the crossed channels just the opposite).
The squared matrix element $| {\cal T}^{(X)}(\pi^{\pm}) | ^2$ in Eq.~(\ref{GX1})
is a combination of contributions from $N_1$ and $N_2$ and
from the two channels $D$ (direct) and $C$ (crossed), and is
given explicitly in Eq.~(\ref{calTX}) in Appendix \ref{app1}.

Combining Eqs.~(\ref{GX1}) and (\ref{calTX}), we obtain
\ba
\Gamma^{(X)}(\pi^{\pm}) &=&
\sum_{i=1}^2 \sum_{j=1}^2 k_{i,\pm}^{(X) *} k_{j,\pm}^{(X)}
{\big [}
\bG^{(X)}(DD^{*})_{ij} + \bG^{(X)}(CC^{*})_{ij}
+ \bG_{\pm}^{(X)}(DC^{*})_{ij} + \bG_{\pm}^{(X)}(CD^{*})_{ij} {\big ]} \ ,
\label{GX2}
\ea
where $X=$LV, LC; 
the indices $(i, j)$ indicate contributions from $N_i$ and $N_j$
neutrino exchange amplitudes; 
and $k_{j,\pm}^{(X)}$ are the corresponding heavy-light mixing factors
\be
\label{kj}
k_{j,+}^{{\rm (LV)}}  =  B_{e N_j}^2 \ , \qquad k_{j,+}^{{\rm (LC)}} = B_{e N_j} B^{*}_{\mu N_j} \ , \qquad k_{j,-}^{(X)} = \left( k_{j,+}^{(X)} \right)^{*} \ .
\ee
In Eq.~(\ref{GX2}) we denoted by $\bG^{(X)}(YZ^{*})_{ij}$ ($i,j=1,2$) 
the elements of the 
normalized (i.e., without mixings) decay width matrices
$\bG^{(X)}(YZ^{*})$ ($X=$LV, LC; $Y, Z = D, C$)
\be
\label{GXij}
\bG_{\pm}^{(X)}(XY^{*})_{ij} = 
K^2 \; \frac{1}{2!} \frac{1}{2 M_{\pi}} \frac{1}{(2 \pi)^8} \int d_4 \;
P_i^{(X)}(Y) P_j^{(X)}(Z)^{*} \; T_{\pm}^{(X)}(YZ^{*}) \ ,
\ee
where the expressions for $T_{\pm}^{(X)}( YZ^{*})$ (with $X=$LV, LC)
for the direct ($YZ^{*}=DD^{*}$), crossed ($YZ^{*}=CC^{*}$) and 
direct-crossed interference ($YZ^{*}=DC^{*}$, $CD^{*}$)
appearing in Eq.~(\ref{GXij}) 
are given in Appendix \ref{app1}, Eqs.~(\ref{TLV})-(\ref{TLC}).
We note that $T_{+}^{(X)}(DD^{*})=T_{-}^{(X)}(DD^{*})$ and 
$T_{+}^{(X)}(CC^{*})=T_{-}^{(X)}(CC^{*})$, so that the terms
$\bG^{(X)}(DD^{*})_{ij}$ and $\bG^{(X)}(CC^{*})_{ij}$ in Eq.~(\ref{GX2}) have
no subscripts $\pm$.
In Eq.~(\ref{GXij}), $P_j^{(X)}(Y)$  ($X=$LV, LC) represent the  $N_j$ propagator functions of the direct and crossed channels ($Y=D, C$)
\bes
\label{Pj}
\ba
P_j^{\rm (LC)}(D) &=& \frac{1}{\left[ (p_{\pi}-p_1)^2 - M_{N_j}^2 + i \Gamma_{N_j} M_{N_j} \right]} , \qquad
P_j^{\rm (LV)}(D) = M_{N_j} P_j^{\rm (LC)}(D) ,
\label{PjD}
\\
P_j^{\rm (LC)}(C) &=& \frac{1}{\left[ (p_{\pi}-p_2)^2 - M_{N_j}^2 + i \Gamma_{N_j} M_{N_j} \right]} , \qquad
P_j^{\rm (LV)}(C) = M_{N_j} P_j^{\rm (LC)}(C)  ,
\label{PjC}
\ea
\ees
and $K^2$ constant is
\be
K^2 = G_F^4 f_{\pi}^2 |V_{ud}|^2 \approx 2.983 \times 10^{-22}   \ {\rm GeV}^{-6} \ .
\label{Ksqr}
\ee
Several symmetry relations are valid between the normalized matrices
(\ref{GXij}), cf.~Eqs.~(\ref{symm}) in Appendix \ref{app1}; the most
important is that $\bG^{(X)}(DD^{*})=\bG^{(X)}(CC^{*})$ and that this
($2 \times 2$) matrix is self-adjoint.
Later we will see that the direct-crossed interference contributions
$\bG^{(X)}(DC^{*}), \bG^{(X)}(CD^{*})$
are suppressed by several orders of magnitude in comparison to
$\bG^{(X)}(DD^{*})$.

The branching ratios are obtained by dividing the
calculated decay widths $\Gamma^{(X)}(\pi^{\pm})$, Eqs.~(\ref{not0})-(\ref{GX1})
and (\ref{GX2}),
by the total decay width of the charged pion $\Gamma(\pi^+ \to {\rm all})$
\be
\Gamma(\pi^+ \to {\rm all}) = 2.529 \times 10^{-17} \ {\rm GeV}
\approx \frac{1}{8\pi} G^{2}_{F}f^{2}_{\pi} M^{2}_{\mu} M_{\pi} |V_{ud}|^2 \left ( 1-\frac{M^{2}_{\mu}}{M^{2}_{\pi}} \right )^2 \ .
\label{Gpiall}
\ee
Another important quantity in the evaluations of  $\bG^{(X)}(YZ^{*})$
[and ${\rm Br}^{(X)}(YZ^{*})$] is the total decay width
$\Gamma_{N_j}$ of the intermediate on-shell neutrinos,
which for the mass range of interest [Eq.~(\ref{MNjint})]
can be approximated in the following way:
\begin{equation}
\Gamma_{N_j} \approx {\cal C}\  \K_j \bG(M_{N_j}) \ ,
\label{DNwidth}
\end{equation}
where
\begin{equation}
 \bG(M_{N_j}) \equiv \frac{G_F^2 M_{N_j}^5}{192\pi^3} \ ,
\label{barG}
\ee
and ${\cal C} =2$ if $N_j$ is Majorana neutrino, and
${\cal C} =1$ if $N_j$ is Dirac neutrino. The factor $\K_j$
includes the heavy-light mixing factors dependence,
from the charged channels
and the neutral interaction channels mediated by $Z$.
Using the results of Appendix C of Ref.~\cite{Atre},
the factor $\K_j$ can be obtained
\begin{equation}
\K_j \approx 1.6 \; |B_{e N_j}|^2 + 1.1 \; ( |B_{\mu N_j}|^2 + |B_{\tau N_j}|^2 ) \ ,
\quad (j=1,2) \ .
\label{calK}
\end{equation}
The charged and neutral channel contributions produce only (light)
neutrinos and $e^+ e^-$; decays with muon in the final state are
suppressed by a kinematical factor $f(M^2_{\mu}/M^2_{N_j}) <10^{-2}$ and are
neglected in the formula (\ref{calK}).\footnote{
Eq.~(\ref{calK}) is obtained by using Eqs.~(C.6)-(C.9) of Ref.~\cite{Atre},
for the channels $N_j \to e^+ e^- \nu_{\ell}$, 
$\nu_{\ell} {\overline \nu}_{\ell} \nu_{\ell'}$. The coefficients
in the corresponding formula (2.3) of Ref.~\cite{JHEP} are not correct.}

\section{The branching ratios and the CP asymmetry for the rare decays}
\label{sec:ACPsum}

In this Section we use the results of the previous Section to obtain
the results for the branching ratios ${\rm Br}_{\pm}^{(X)}$
and the CP asymmetry ratios ${\cal A}^{(X)}_{\rm CP}$ ($X=LP,LC$)
of the discussed rare processes
\ba
{\rm Br}_{\pm}^{(X)} &=& \frac{S^{(X)}_{\pm}(\pi)}{\Gamma(\pi^+ \to {\rm all})} \equiv  \frac{ \Gamma^{(X)}(\pi^-) \pm \Gamma^{(X)}(\pi^+)}{\Gamma(\pi^- \to {\rm all})} \ ,
\label{Brdef}
\\
{\cal A}^{(X)}_{\rm CP} &=& \frac{{\rm Br}_{-}^{(X)}}{{\rm Br}_{+}^{(X)}}=
 \equiv \frac{ \Gamma^{(X)}(\pi^-) - \Gamma^{(X)}(\pi^+)}{ \Gamma^{(X)}(\pi^-) + \Gamma^{(X)}(\pi^+)} \ ,
\label{Adef}
\ea
where we recall the use of notations (\ref{not0}).
The total branching ratios are
${\rm Br}_{\pm} ={\rm Br}_{\pm}^{\rm (LV)}+{\rm Br}_{\pm}^{\rm (LC)}$ 
when $N_j$ are Majorana neutrinos, and ${\rm Br}_{\pm} ={\rm Br}_{\pm}^{\rm (LC)}$
when $N_j$ are Dirac neutrinos.
It is useful to introduce the following notations related
with the heavy-light neutrino mixing elements $B_{e N_j}$ and $B_{\mu N_j}$,
where we adopt the convention $M_{N_2} > M_{N_1}$:
\bes
\label{not}
\ba
\kappa_e & = & \frac{|B_{e N_2}|}{|B_{e N_1}|} \ ,
\quad
\kappa_{\mu} =  \frac{|B_{\mu N_2}|}{|B_{\mu N_1}|} \ ,
\label{kap}
\\
B_{e N_j} &=& |B_{e N_j}| e^{i \theta_{e j}} \ ,
\quad
B_{\mu N_j} = |B_{\mu N_j}| e^{i \theta_{\mu j}} \ ,
\label{thellj}
\\
\theta^{\rm (LV)} & = & 2 (\theta_{e 2} - \theta_{e 1}) \ ,
\quad
\theta^{\rm (LC)} =  (\theta_{e 2} - \theta_{e 1}) -
 (\theta_{\mu 2} - \theta_{\mu 1}) \ .
\label{delth}
\ea
\ees
It turns out (see later) that in our cases of interest the $D$-$C$ interference
contributions are negligible, and the resulting (sums) $S_{+}^{(X)}(\pi)$
of the decay widths are
\bes
\label{SplX}
\ba
S_{+}^{\rm (LV)}(\pi) & \equiv & \left( \Gamma^{\rm (LV)}(\pi^-) + \Gamma^{\rm (LV)}(\pi^+) \right)
\nonumber\\
&= &
4 |B_{e N_1}|^4 \bG^{\rm (LV)}(DD^{*})_{11}
\left[ 1 +
\kappa_e^4 \frac{\bG^{\rm (LV)}(DD^{*})_{22}}{\bG^{\rm (LV)}(DD^{*})_{11}}
+ 2 \kappa_e^2 \left( \cos \theta^{\rm (LV)} \right)
\delta_1^{\rm (LV)} \right] \ ,
\label{SplLV}
\\
S_{+}^{\rm (LC)}(\pi) & \equiv & \left( \Gamma^{\rm (LC)}(\pi^-) + \Gamma^{\rm (LC)}(\pi^+) \right)
\nonumber\\
&= &
4 |B_{e N_1}|^2 |B_{\mu N_1}|^2  \bG^{\rm (LC)}(DD^{*})_{11}
\left[ 1 +
\kappa_e^2 \kappa_{\mu}^2 \frac{\bG^{\rm (LC)}(DD^{*})_{22}}{\bG^{\rm (LC)}(DD^{*})_{11}}
+ 2 \kappa_e \kappa_{\mu}  \left( \cos \theta^{\rm (LC)} \right)
\delta_1^{\rm (LC)} \right] \ ,
\label{SplLC}
\ea
\ees
where $\delta_j^{(X)}$ in the above quantities represent the
(relative) contribution of the $N_1$-$N_2$ interference channel
\be
\delta_j^{(X)} \equiv \frac{{\rm Re} \bG^{(X)}(DD^{*})_{12}}{ \bG^{(X)}(DD^{*})_{jj}} \ , \quad (X=LV, LC; \; j=1,2) \ .
\label{delX}
\ee
On the other hand, the difference $S_{-}^{(X)}(\pi)$ of the $\pi^{-}$ and $\pi^{+}$
rare decays is (where the $D$-$C$ interference terms are neglected)
\bes
\label{SmiX}
\ba
S_{-}^{\rm (LV)}(\pi) & \equiv & \left( \Gamma^{\rm (LV)}(\pi^-) - \Gamma^{\rm (LV)}(\pi^+) \right)
=
8 |B_{e N_1}|^4 \kappa_e^2   \left( \sin \theta^{\rm (LV)} \right)
{\rm Im} \bG^{\rm (LV)}(DD^{*})_{12} \ ,
\label{SmiLV}
\\
S_{-}^{\rm (LC)}(\pi) & \equiv &
\left( \Gamma^{\rm (LC)}(\pi^-) - \Gamma^{\rm (LC)}(\pi^+) \right)
=
8 |B_{e N_1}|^2 |B_{\mu N_1}|^2 \kappa_e \kappa_{\mu}
\left( \sin \theta^{\rm (LC)} \right)
{\rm Im} \bG^{\rm (LC)}(DD^{*})_{12} \ .
\label{SmiLC}
\ea
\ees
In these expressions we can recognize (a posteriori) the difference of
the CP-odd phases as $\theta^{(X)}$ ($X=LV, LC$) coming from the
PMNS mixing matrix elements, cf.~Eqs.~(\ref{thellj})-(\ref{delth});
while (sinus of) the difference of the CP-even phases is contained
in the imaginary part of the product of propagators,
${\rm Im} \bG^{(X)}(DD^{*})_{12} \propto  {\rm Im} P_1^{(X)}(D) P_2^{(X)}(D)^{*}$,
cf.~Eqs.~(\ref{ImP1P2gen}) later.

In the limit of $\Gamma_{N_j} \to +0$, i.e., $\Gamma_{N_j} \ll M_{N_j}$,
the expression for the ``diagonal'' decay
width $\bG^{(X)}(DD^{*})_{11}$ [and thus also for $\bG^{(X)}((DD^{*})_{22}$] can
be calculated analytically. The differential decay width $d \Gamma^{(X)}/d E_{\mu}$
with respect to the muon energy $E_{\mu}$, in the $N_j$ rest frame, was
obtained in Ref.~\cite{JHEP}, and the result of explicit integration of it
over $E_{\mu}$, for the general case of not neglected electron mass
($M_e \not= 0$), is
\ba
\bG^{(X)}(DD^{*})_{jj} & \equiv & \bG(DD^{*})_{jj}
= \frac{K^2}{192 (2 \pi)^4} \frac{M_{N_j}^{11}}
{ M_{\pi}^3 \Gamma_{N_j} } \lambda^{1/2}(x_{\pi j}, 1, x_{e j})
\left[ x_{\pi j} -1
+ x_{e j}(x_{\pi j} + 2 - x_{e j}) \right] {\cal F}(x_j, x_{e j}) \ ,
\label{GXDDp}
\ea
where we use the notations
\bes
\label{notGXDDp}
\ba
\lambda(y_1,y_2,y_3) & = & y_1^2 + y_2^2 + y_3^2 - 2 y_1 y_2 - 2 y_2 y_3 - 2 y_3 y_1 \ ,
\label{lambda}
\\
x_{\pi j} &=& \frac{M_{\pi}^2}{M_{N_j}^2} \ , \quad
x_{ej} =  \frac{M_e^2}{M_{N_j}^2} \ , \quad
x_j =\frac{M_{\mu}^2}{M_{N_j}^2} \ ,  \quad (j=1,2) \ ,
\label{xjs}
\ea
\ees
and the function ${\cal F}(x_j, x_{e j})$ is given in Appendix \ref{app2}
[Eq.~(\ref{calF})] where the derivation of this expression (\ref{GXDDp})
is given.
When $M_e=0$, the results acquires a simpler form
\be
\lim_{M_e \to 0} \bG^{(X)}(DD^{*})_{jj} = \frac{K^2}{192 (2 \pi)^4} \frac{M_{N_j}^{11}}
{\Gamma_{N_j} M_{\pi}^3} ( x_{\pi j} - 1 )^2
f (x_j) \ ,
\label{GXDDp0}
\ee
where the function $f(x_j) = {\cal F}(x_j,0)$ is
\be
f(x_j) = 1 - 8 x_j + 8 x_j^3 - x_j^4 - 12 x_j^2 \ln x_j \ .
\label{fx}
\ee
We note that the expression (\ref{GXDDp}) is the same
for $X=LV$ and $X=LC$. In the range of the masses
$0.117 \ {\rm GeV} < M_{N_j} < 0.136 \ {\rm GeV}$ the expression (\ref{GXDDp0})
differs from the exact expression (\ref{GXDDp}) [with Eq.~(\ref{calF})]
by less than one per cent.
However, for $0.106 \ {\rm GeV} < M_{N_j} < 0.117 \ {\rm GeV}$ and for
$0.136 \ {\rm GeV} < M_{N_j} < 0.139 \ {\rm GeV}$ the deviation is more than
one per cent. For values of $M_{N_j}$ close to the lower on-shell bound
$M_{\mu}+M_e$ ($ \approx 0.1062$ GeV) the deviation is very large and
the expression (\ref{GXDDp}) [with Eq.~(\ref{calF})] must be used instead
of Eq.~(\ref{GXDDp0}) for $\bG^{(X)}(DD^{*})_{jj}$. We will use the general
expression (\ref{GXDDp}) unless otherwise stated.

Furthermore, we can also calculate analogously as $\bG^{(X)}(DD^{*})_{jj}$
the analytic expression for the asymmetric difference $S_{-}^{(X)}$
in the limit $\Gamma_{N_j} \to +0$ ($\Gamma_{N_j} \ll M_{N_2}-M_{N_1}$).
In order to explain this analogy, we note
that in the limit $\Gamma_{N_j} \to +0$ it was crucial to use
in the analytic calculation of  $\bG^{(X)}(DD^{*})_{jj}$ the
identity
\ba
|P_j^{\rm (LC)}(D)|^2 &=&
\left | \frac{1}{(p_{\pi}-p_1)^2-M^{2}_{N_j}+i \Gamma_{N_j} M_{N_j}} \right | ^2
\nonumber\\
&\approx &
\frac{\pi}{M_{N_j} \Gamma_{N_j}} \delta((p_{\pi}-p_1)^2-M^{2}_{N_j})\ ;
\quad ( j=1,2; \; \Gamma_{N_j} \ll M_{N_j} ) \ .
\label{P1P1}
\ea
On the other hand, in the difference $S_{-}^{(X)} \propto
{\rm Im} \bG^{(X)}(DD^{*})_{12}$ we have in the integrand of
${\rm Im} \bG^{(X)}(DD^{*})_{12}$ as a factor the
following combination of propagators:
\bes
\label{ImP1P2gen}
\ba
{\rm Im} P_1^{\rm (LC)}(D) P_2^{\rm (LC)}(D)^{*}
&= &
\frac{
\left( p_N^2 - M_{N_1}^2 \right)  \Gamma_{N_2} M_{N_2}
- \Gamma_{N_1} M_{N_1} \left( p_N^2 - M_{N_2}^2 \right)
}
{
\left[ \left( p_N^2 - M_{N_1}^2 \right)^2 + \Gamma_{N_1}^2 M_{N_1}^2
\right]
\left[ \left( p_N^2 - M_{N_2}^2 \right)^2 + \Gamma_{N_2}^2 M_{N_2}^2
\right]
}
\label{ImP1P2ex}
\\
& \approx &
\mathcal{P} \left ( \frac{1}{p^{2}_{N}-M^{2}_{N1}} \right )
\pi\ \delta (p^{2}_{N}-M^{2}_{N2})
-
\pi\ \delta (p^{2}_{N}-M^{2}_{N1})
\mathcal{P} \left ( \frac{1}{p^{2}_{N}-M^{2}_{N2}} \right )
\label{ImP1P2a}
\\
& = & \frac{\pi}{M^{2}_{N2}-M^{2}_{N1}} \left [ \delta  ( p^{2}_{N}-M^{2}_{N2})+ \delta  ( p^{2}_{N}-M^{2}_{N1})  \right ] \ ,
\label{ImP1P2}
\ea
\ees
where $p_N=(p_{\pi}-p_1)$ in the direct channel, and we assumed that
$\Gamma_{N_j} \ll | \Delta M_N | \equiv M_{N_2}-M_{N_1}$ in Eqs.~(\ref{ImP1P2a})-(\ref{ImP1P2}).
When $X=LV$, the
corresponding combination of propagators is the same as in
Eq.~(\ref{ImP1P2}) but with the additional factor $M_{N_1} M_{N_2}$.
The expression (\ref{ImP1P2})
has formally the same structure as the expression (\ref{P1P1}), except
for the factors in front of the delta(s). Therefore, the integration over
the final phase space can be performed formally in the same way. This
then results in the expressions
\bes
\label{ImG12}
\ba
\lefteqn{
{\rm Im} \bG^{\rm (LV)}(DD^{*})_{12} =
\eta^{\rm (LV)} \; \frac{K^2}{192 (2 \pi)^4} \frac{1}{M_{\pi}^3}
 \frac{M_{N_1} M_{N_2}}{(M_{N_2}+M_{N_1}) (M_{N_2}-M_{N_1})}
}
\nonumber\\
&& \times
\sum_{j=1}^2 M_{N_j}^{10}  \lambda^{1/2}(x_{\pi j}, 1, x_{e j})
\left[ x_{\pi j} -1 + x_{e j}(x_{\pi j} + 2 - x_{e j}) \right]
{\cal F}(x_j, x_{e j}) \ ,
\label{ImG12LV}
\\
\lefteqn{
{\rm Im} \bG^{\rm (LC)}(DD^{*})_{12} =
\eta^{\rm (LC)} \; \frac{K^2}{192 (2 \pi)^4} \frac{1}{M_{\pi}^3}
\frac{1}{(M_{N_2}+M_{N_1})(M_{N_2}-M_{N_1})}
}
\nonumber\\
&& \times
\sum_{j=1}^2 M_{N_j}^{12}
 \lambda^{1/2}(x_{\pi j}, 1, x_{e j})
\left[ x_{\pi j} -1 + x_{e j}(x_{\pi j} + 2 - x_{e j}) \right]
{\cal F}(x_j, x_{e j}) \ ,
\label{ImG12LC}
\ea
\ees
where the overall factor $\eta^{(X)}$ is equal to unity ($\eta^{(X)} = 1$) when
$\Gamma_{N_j} \ll | \Delta M_N |$, i.e., when the identity (\ref{ImP1P2})
can be applied. Nonetheless, when $\Gamma_{N_j} \not\ll | \Delta M_N |$,
we have in general corrections to these formulas, in the form of
$\eta < 1$,\footnote{
We note that there is no such overall correction factor in the
expression (\ref{GXDDp}) for $\bG^{(X)}(DD^{*})_{jj}$, because in
our considered cases $\Gamma_{N_j} \ll M_{N_j}$ always and Eq.~(\ref{GXDDp})
is the correct expression then.}
and the exact expression (\ref{ImP1P2ex}) has to be used instead of
the approximation (\ref{ImP1P2}).

All these quantities can be evaluated also via numerical integrations
over the final phase space, with finite widths $\Gamma_{N_j}$ in the
propagators. The scalings
$\bG^{(X)}(DD^{*})_{jj} \propto \Gamma_{N_j}$,
${\rm Im}  \bG^{(X)}(DD^{*})_{12}$ $\propto 1/\Delta M_N$,
as suggested
by Eqs.~(\ref{GXDDp}) and  (\ref{ImG12}), are confirmed numerically
(when $\Gamma_{N_j} \ll M_{N_j}$, and $\Gamma_{N_j} \ll \Delta M_N$, respectively).
Furthermore, the numerical evaluations indicate clearly that the
direct-crossed ($DC^{*}$ and $CD^{*}$)
interference contributions to $S_{\pm}^{(X)}(\pi)$
are negligible in all considered cases, in comparison with the corresponding
direct ($DD^{*}$) and crossed channel ($CC^{*}$) contributions.
Namely, in the sum $S_{+}^{(X)}(\pi)$,
the interference contributions ${\rm Re} \bG^{(X)}(DC^{*})_{ij}
\sim 10^{-37} \ {\rm GeV}$
are approximately independent of $\Gamma_{N_j}$.
On the other hand,
$\bG^{(X)}(DD^{*})_{jj}=\bG^{(X)}(CC^{*})_{jj}$
is at $\Gamma_N=10^{-4}$ GeV about two orders of magnitude
larger than ${\rm Re} \bG^{(X)}(DC^{*})_{ij}$.
$\bG^{(X)}(DD^{*})_{jj}$ grows at decreasing $\Gamma_N$ as
$1/\Gamma_N$ [Eq.~(\ref{GXDDp})],
while ${\rm Re} \bG^{(X)}(DC^{*})_{ij}$ does not increase and becomes thus at
$\Gamma_N < 10^{-4}$ GeV relatively negligible.

In the difference (asymmetry) $S_{-}^{(X)}(\pi)$, the $DC^{*}$ interference
contribution
${\rm Im}  \bG^{(X)}(DC^{*})_{12} \sim 10^{-38} \ {\rm GeV}$ is approximately
independent of $\Delta M_N$.
On the other hand, ${\rm Im}  \bG^{(X)}(DD^{*})_{12} = {\rm Im}  \bG^{(X)}(CC^{*})_{12}$
is at $\Delta M_N = 10^{-3}$ GeV about two orders of magnitude
larger than ${\rm Im}  \bG^{(X)}(DC^{*})_{12}$.
${\rm Im}  \bG^{(X)}(DD^{*})_{12}$  grows at decreasing  $\Delta M_N$ as
$1/\Delta M_N$ [Eq.~(\ref{ImG12})],
while  ${\rm Im}  \bG^{(X)}(DC^{*})_{12}$ does not increase and becomes thus at
$\Delta M_N < 10^{-3}$ GeV relatively negligible.

On the other hand, the numerical evaluations with $\Gamma_{N_j} \not \ll \Delta M_N$
give us the values of the $\delta_j^{(X)}$ [cf.~Eqs.~(\ref{delX}) and (\ref{SplX})]
and $\eta^{(X)}$ correction terms,
due to non-negligible overlap of the $N_1$ with $N_2$ resonance.
It turns out that these functions are independent of $X$ ($=LV,LC$),
and that $\eta$ and $\delta \equiv (1/2)(\delta_1 + \delta_2)$
are effectively functions of only one
parameter, $y \equiv \Delta M_N/\Gamma_N$, where $\Delta M_N \equiv M_{N_2}-M_{N_1}$
($> 0$), and $\Gamma_N = (1/2)(\Gamma_{N_1} +\Gamma_{N_2})$
\bes
\label{etadel}
\ba
\eta &=& \eta(y) \ , \quad y \equiv \frac{\Delta M_N}{\Gamma_N} \ ,
\quad \Gamma_N \equiv \frac{1}{2} (\Gamma_{N_1} + \Gamma_{N_2}) \ ,
\label{etadel1}
\\
\delta &=&\delta(y) \ , \quad \delta \equiv  \frac{1}{2} (\delta_1 + \delta_2) \ ,
\quad
\frac{\delta_1}{\delta_2} = \frac{\bG(DD^*)_{22}}{\bG(DD^*)_{11}} =
\frac{\Gamma_{N_1}}{\Gamma_{N_2}} = \frac{\K_1}{\K_2} \ .
\label{etadel2}
\ea
\ees
The values of $\delta$ ($=\delta^{(X)}$) and $\eta$ ($=\eta^{(X)}$) as functions
of $\Delta M_N/\Gamma_N$ can be obtained by numerical integrations over the
four-particle finite phase space, and are tabulated in Table \ref{tabdelet}
(with their estimated uncertainties due to numerical integrations).
\begin{table}
\caption{Values of $\delta(y)$ correction terms and $\eta(y)/y$
correction factors for various values of $y \equiv  \Delta M_N/\Gamma_N$.}
\label{tabdelet}
\begin{tabular}{ll|lll}
$y \equiv \frac{\Delta M_N}{\Gamma_N}$ & $\log_{10} y$ &
$\delta(y)$ & $\eta(y)$ & $\frac{\eta(y)}{y}$
\\
\hline
10.0 & 1.000 & $0.0100 \pm 0.0005$ & $0.984 \pm 0.003$ & $0.0984 \pm 3 \times 10^{-4}$
\\
5.00 & 0.699 & $0.038 \pm 0.002$ &$ 0.957 \pm 0.003$ & $0.191 \pm 0.001$
\\
2.50 & 0.398 & $0.137 \pm 0.006$ & $0.854 \pm 0.003$ & $0.342 \pm  0.001$
\\
1.67 & 0.222 & $0.265 \pm 0.005$ & $0.730 \pm 0.005$ & $0.438 \pm 0.003$
\\
1.25 & 0.097 & $0.392 \pm 0.006$ & $0.610 \pm 0.007$ & $0.488 \pm 0.006$
\\
1.00 & 0.000 & $0.505 \pm 0.010$ & $0.498 \pm 0.005$ & $0.498 \pm 0.005$
\end{tabular}
\end{table}

We note that the rare process decay widths $S_{+}^{(X)}(\pi)$, Eq.~(\ref{SplX}),
are formally quartic in the heavy-light
mixing elements $|B_{\ell N}|$, i.e., very small.
Nonetheless, they are proportional to the expressions $\bG(DD^{*})_{jj}$,
Eq.~(\ref{GXDDp}), which in turn is proportional to $1/\Gamma_{N_j}$
due to the on-shellness of the intermediate $N_j$'s. This $1/\Gamma_{N_j}$
is proportional to $1/\K_j \sim 1/|B_{\ell N_j}|^2$ according to
Eqs.~(\ref{DNwidth})-(\ref{calK}). Therefore, the on-shellness of $N_j$'s
makes the rare process decay widths significantly less suppressed by
the mixings:
\bes
\label{onsh}
\ba
\bG(DD^{*})_{jj} &\propto& 1/\Gamma_{N_j} \propto 1/\K_j
\propto 1/|B_{\ell N_j}|^2 \ ,
\\
S_{+}^{(X)}(\pi) & \propto & |B_{\ell N_j}|^2 \ .
\ea
\ees
On the other hand, comparing the expressions (\ref{ImG12})
relevant for the CP asymmetries $S_{-}^{(X)}(\pi)$ (\ref{SmiX}),
with the expression (\ref{GXDDp}) relevant for the decay widths
$S_{+}^{(X)}(\pi)$ (\ref{SplX}), we see that the asymmetries $S_{-}^{(X)}(\pi)$ are
suppressed by mixings as $\sim |B_{\ell N}|^4$, making them in general
much smaller than the decay widths $S_{+}^{(X)}(\pi) \propto |B_{\ell N_j}|^2$.
However, the asymmetries are proportional to $1/\Delta M_N$ (where
$\Delta M_N = M_{N_2} - M_{N_1} > 0$), cf.~Eqs.~(\ref{ImG12}).
In general, $\Delta M_N \gg \Gamma_{N_j}$.
Nonetheless, in a scenario where $\Delta M_N$ becomes very small and (almost)
comparable with $\Gamma_{N_j}$, the asymmetries $S_{-}^{(X)}(\pi)$ can become
comparable with the decay widths $S_{+}^{(X)}(\pi)$. A model with two
almost degenerate neutrinos $N_j$ in the mass range of $\sim 10^2$ eV
has been constructed and investigated in Ref.~\cite{Shapo}.

In particular, in this limit of two almost degenerate neutrinos $N_j$,
where now $M_{N_1} \approx M_{N_2} \equiv M_N$,
the formulas (\ref{GXDDp}), (\ref{delX}) and (\ref{ImG12}) get simplified.
In this case, it is convenient to introduce a ``normalized'' branching ratio $\Br$
\ba
{\overline {\rm Br}}(M_N) & \equiv &
\frac{1}{4 \pi} \frac{K^2 M_{\pi}^3}{G_F^2 \Gamma(\pi^+ \to {\rm all})}
\frac{1}{x_{\pi}^3} \lambda^{1/2}(x_{\pi}, 1, x_{e})
\left[ x_{\pi} -1 + x_{e}(x_{\pi} + 2 - x_{e}) \right]
{\cal F}(x, x_{e}) \ ,
\label{bBr}
\ea
where we use the notations
\ba
x_{\pi} &=& \frac{M_{\pi}^2}{M_{N}^2} \ , \quad
x_{e} =  \frac{M_e^2}{M_{N}^2} \ , \quad
x =\frac{M_{\mu}^2}{M_{N}^2} \ .
\label{xs}
\ea
In terms of this branching ratio ${\overline {\rm Br}}$, 
the formulas (\ref{GXDDp}), (\ref{delX})
and (\ref{ImG12}) can be rewritten, in the mentioned almost degenerate scenario,
as
\bes
\label{Gnor}
\ba
\frac{\bG(DD^{*})_{jj}}{\Gamma(\pi^+ \to {\rm all})} &=&
\frac{1}{4 {\cal C} \K_j} \Br \ ,
\label{Gjjnor}
\\
\frac{{\rm Re} \bG(DD^{*})_{12}}{\Gamma(\pi^+ \to {\rm all})} &=&
\frac{\delta(y)}{2 {\cal C} (\K_1+\K_2)} \Br \ ,
\label{ReG12nor}
\\
\frac{{\rm Im} \bG(DD^{*})_{12}}{\Gamma(\pi^+ \to {\rm all})} &=&
\frac{\eta(y)/y}{2 {\cal C} (\K_1+\K_2)} \Br \ ,
\label{ImG12nor}
\ea
\ees
where $y \equiv \Delta M_N/\Gamma_N$.
Similarly, after some algebra,
we can rewrite in this scenario ($M_{N_1} \approx M_{N_2} \equiv M_N$)
the obtained branching ratios ${\rm Br}_{\pm}$ 
and CP asymmetry ratios ${\cal A}_{\rm CP}$ for the considered
rare decays, in terms of
$\Br$ and of the heavy-light mixing parameters. Below we present the
results for the case when the neutrinos $N_j$ are Dirac (Di), and when they
are Majorana (Ma) neutrinos. The branching ratio ${\rm Br}_{+}$
for the considered rare processes is
\bes
\label{Brpl}
\ba
{\rm Br}_{+}^{\rm (Di)} & \equiv &
\frac{S_{+}^{\rm (LC)}(\pi)}{\Gamma(\pi^+ \to {\rm all})}
\nonumber\\
&=&  {\bigg [} \sum_{j=1}^2 \frac{|B_{e N_j}|^2 |B_{\mu N_j}|^2}{\K_j}
+ 4 \delta(y) \frac{|B_{e N_1}||B_{e N_2}||B_{\mu N_1}||B_{\mu N_2}|}{(\K_1+\K_2)}
\cos \theta^{\rm (LC)} {\bigg ]} \Br(M_N)
\label{BrplDi1}
\\
&=&  \frac{|B_{e N_1}|^2 |B_{\mu N_1}|^2}{\K_1} \left[
1\! + \! \frac{\K_1}{\K_2} \kappa_e^2 \kappa_{\mu}^2 \! + \!
4 \delta(y) \frac{\K_1}{(\K_1+\K_2)} \kappa_e^2 \kappa_{\mu}^2
\cos \theta^{\rm (LC)} \right] \Br(M_N) \ ,
\label{BrplDi2}
\\
{\rm Br}_{+}^{\rm (Ma)} & \equiv &
\frac{S_{+}^{\rm (LV)}(\pi)+S_{+}^{\rm (LC)}(\pi)}{\Gamma(\pi^+ \to {\rm all})}
\nonumber\\
&=&  {\bigg [} \sum_{j=1}^2
\frac{|B_{e N_j}|^2 ( |B_{e N_j}|^2 + |B_{\mu N_j}|^2)}{2 \K_j}
\nonumber\\
&&+ 2 \delta(y) \frac{|B_{e N_1}||B_{e N_2}|}{(\K_1+\K_2)}
\left( |B_{e N_1}||B_{e N_2}| \cos \theta^{\rm (LV)} +
 |B_{\mu N_1}||B_{\mu N_2}| \cos \theta^{\rm (LC)} \right) {\bigg ]} \Br(M_N)
\label{BrplMa1}
\\
&=&  \frac{|B_{e N_1}|^2 (|B_{e N_1}|^2+|B_{\mu N_1}|^2)}{2 \K_1} {\bigg [}
1 + \frac{\K_1}{\K_2} \kappa_e^2 \left(
\frac{\kappa_e^2 |B_{e N_1}|^2 +  \kappa_{\mu}^2 |B_{\mu N_1}|^2}{|B_{e N_1}|^2+|B_{\mu N_1}|^2} \right)
\nonumber\\
&&+ 4 \delta(y) \frac{\K_1}{(\K_1+\K_2)} \kappa_e
{\bigg (}
\frac{ \kappa_e |B_{e N_1}|^2}{(|B_{e N_1}|^2+|B_{\mu N_1}|^2)}
\cos \theta^{\rm (LV)}
+
\frac{\kappa_{\mu} |B_{\mu N_1}|^2}{(|B_{e N_1}|^2+|B_{\mu N_1}|^2)}
\cos \theta^{\rm (LC)}
{\bigg )} {\bigg ]}  \Br(M_N) .
\label{BrplMa2}
\ea
\ees
Here we took into account that in the Dirac case only the
$LC$ process contributes, while in the Majorana case both
the lepton number violating ($LV$) and conserving ($LC$) processes
contribute. The mixing parameters $\K_j$ ($\sim |B_{\ell N_j}|^2$)
are given in
Eq.~(\ref{calK}), and we took into account that in
Eq.~(\ref{DNwidth}) for $\Gamma_{N_j}$ the factor ${\cal C}$ is one in
the Dirac case and is two in the Majorana case. The contributions
of the $N_1$-$N_2$ overlap effects give the relative corrections of
${\cal O}(\delta)$ and are negligible when $\Delta M_N > 10 \Gamma_N$,
cf.~Table \ref{tabdelet}.

The (CP asymmetry) branching ratio ${\rm Br}_{-}$ for the considered rare processes is
\bes
\label{Brmi}
\ba
{\rm Br}_{-}^{\rm (Di)} &\equiv& \frac{S_{-}^{\rm (LC)}(\pi)}{\Gamma(\pi^+ \to {\rm all})}
= \frac{\Gamma^{\rm (LC)}(\pi^-)-\Gamma^{\rm (LC)}(\pi^+)}{\Gamma(\pi^+ \to {\rm all})}
\nonumber\\
&=& \frac{4 |B_{e N_1}| |B_{e N_2}| |B_{\mu N_1}|  |B_{\mu N_2}|}{(\K_1+\K_2)}
\sin \theta^{\rm (LC)} \frac{\eta(y)}{y} \Br(M_N)
\label{BrmiDi1}
\\
&=& \frac{4 |B_{e N_1}|^2 |B_{e N_2}| ^2 \kappa_e \kappa_{\mu}}{(\K_1+\K_2)}
\sin \theta^{\rm (LC)}  \frac{\eta(y)}{y} \Br(M_N)
\label{BrmiDi2}
\\
{\rm Br}_{-}^{\rm (Ma)} &\equiv&
\frac{(S_{-}^{\rm (LV)}(\pi)+S_{-}^{\rm (LC)}(\pi))}{\Gamma(\pi^+ \to {\rm all})}
= \frac{\Gamma^{\rm (LV)}(\pi^-)+\Gamma^{\rm (LC)}(\pi^-)
-\Gamma^{\rm (LV)}(\pi^+)-\Gamma^{\rm (LC)}(\pi^+)}{\Gamma(\pi^+ \to {\rm all})}
\nonumber\\
&=&
\frac{2 |B_{e N_1}| |B_{e N_2}|}{(\K_1+\K_2)}
\left(  |B_{e N_1}| |B_{e N_2}| \sin \theta^{\rm (LV)} +
|B_{\mu N_1}| |B_{\mu N_2}| \sin \theta^{\rm (LC)} \right)
\frac{\eta(y)}{y} \Br(M_N)
\label{BrmiMa1}
\\
&=& \frac{2 \kappa_e |B_{e N_1}|^2}{(\K_1 \! + \! \K_2)} \!
\left(  \kappa_e |B_{e N_1}|^2 \sin \theta^{\rm (LV)} \!\! + \!
\kappa_{\mu} |B_{\mu N_1}|^2 \sin \theta^{\rm (LC)} \right)
\frac{\eta(y)}{y} \Br(M_N) .
\label{BrmiMa2}
\ea
\ees
Consequently, the usual CP asymmetry ratios ${\cal A}_{\rm CP}^{(X)}$ are
obtained from Eqs.~(\ref{Brpl})-(\ref{Brmi})
\bes
\label{ACP}
\ba
\lefteqn{
{\cal A}_{\rm CP}^{\rm (Di)}  \equiv   \frac{{\rm Br}_{-}^{\rm (Di)}}{{\rm Br}_{+}^{\rm (Di)}} =  \frac{\Gamma^{\rm (LC)}(\pi^-)-\Gamma^{\rm (LC)}(\pi^+)}{\Gamma^{\rm (LC)}(\pi^-)+\Gamma^{\rm (LC)}(\pi^+)}
}
\nonumber\\
&=& \frac{ \sin \theta^{\rm (LC)}}{\left[
\frac{1}{4} \frac{|B_{e N_1}|}{|B_{e N_2}|}\frac{|B_{\mu N_1}|}{|B_{\mu N_2}|}
\left(1 + \frac{\K_2}{\K_1} \right) +
\frac{1}{4} \frac{|B_{e N_2}|}{|B_{e N_1}|}\frac{|B_{\mu N_2}|}{|B_{\mu N_1}|}
\left(1 + \frac{\K_1}{\K_2} \right) + \delta(y) \cos \theta^{\rm (LC)}
\right]
} \; \frac{\eta(y)}{y} \ ,
\label{ACPDi}
\\
\lefteqn{
{\cal A}_{\rm CP}^{\rm (Ma)} \equiv  \frac{{\rm Br}_{-}^{\rm (Ma)}}{{\rm Br}_{+}^{\rm (Ma)}} =  \frac{\Gamma^{\rm (LV)}(\pi^-)+\Gamma^{\rm (LC)}(\pi^-)-\Gamma^{\rm (LV)}(\pi^+)-\Gamma^{\rm (LC)}(\pi^+)}{\Gamma^{\rm (LV)}(\pi^-)+\Gamma^{\rm (LC)}(\pi^-)+\Gamma^{\rm (LV)}(\pi^+)+\Gamma^{\rm (LC)}(\pi^+)}
}
\nonumber\\
&=& \frac{ \left( \sin \theta^{\rm (LV)} +  \frac{|B_{\mu N_1}||B_{\mu N_2}|}{|B_{e N_1}||B_{e N_2}|}\sin \theta^{\rm (LC)} \right) }
{\left[
\frac{1}{4} \frac{(|B_{e N_1}|^2 +|B_{\mu N_1}|^2) }{|B_{e N_2}|^2}
\left(1 + \frac{\K_2}{\K_1} \right) +
\frac{1}{4} \frac{(|B_{e N_2}|^2 +|B_{\mu N_2}|^2) }{|B_{e N_1}|^2}
\left(1 + \frac{\K_1}{\K_2} \right) + \delta(y) \left(
\cos \theta^{\rm (LV)} + 
\frac{|B_{\mu N_1}||B_{\mu N_2}|}{|B_{e N_1}||B_{e N_2}|} \cos \theta^{\rm (LC)}
\right)
\right]
} 
\nonumber\\
&&
\times \frac{\eta(y)}{y} \ .
\label{ACPMa}
\ea
\ees
When $y$ ($\equiv \Delta M_N/\Gamma_N$) becomes large ($y > 10$), i.e., when
$\Delta M_N > 10 \Gamma_N$, 
Table \ref{tabdelet} implies that
the CP asymmetries (\ref{Brmi})-(\ref{ACP}) become suppressed by the
$\eta(y)/y$ factor. On the other hand,
when $y < 10$ (i.e., $\Delta M_N < 10 \Gamma_N$)
and $|\theta^{(X)}| \sim 1$, the factor $\eta(y)/y$ is $\sim 1$ and
the CP asymmetry ratios ${\cal A}_{\rm CP}^{(X)}$ become $\sim 1$,\footnote{
If we also assume that
$|B_{\ell N_2}| \approx |B_{\ell N_1}|$ (for $\ell = e, \mu, \tau$),
then also $\K_1 \approx \K_2 \equiv \K$,
and the expressions for ${\cal A}_{\rm CP}$ become particularly simple
\begin{displaymath}
{\cal A}_{\rm CP}^{\rm (Di)} = \frac{\sin \theta^{\rm (LC)}}
{\left(1 + \delta(y) \cos \theta^{\rm (LC)} \right)} \; \frac{\eta(y)}{y}
= \sin \theta^{\rm (LC)} \; \frac{\eta(y)}{y} \left(1 + {\cal O}(\delta) \right) \ ,
\label{ABrkap1Di}
\end{displaymath}
\begin{displaymath}
{\cal A}_{\rm CP}^{\rm (Ma)} =
\left( \frac{|B_{e N_1}|^2 \sin \theta^{\rm (LV)} +
|B_{\mu N_1}|^2 \sin \theta^{\rm (LC)}}{|B_{e N_1}|^2+|B_{\mu N_1}|^2}
\right) \frac{\eta(y)}{y}
\left(1 + {\cal O}(\delta) \right) \ .
\label{ABrkap1Ma}
\end{displaymath}
}
while all ${\rm Br}_{\pm}$ become 
$\sim |B_{\ell N_j}|^2 \Br(M_N)$ ($\ell=e, \mu$).

\begin{figure}[htb]
\centering\includegraphics[width=100mm]{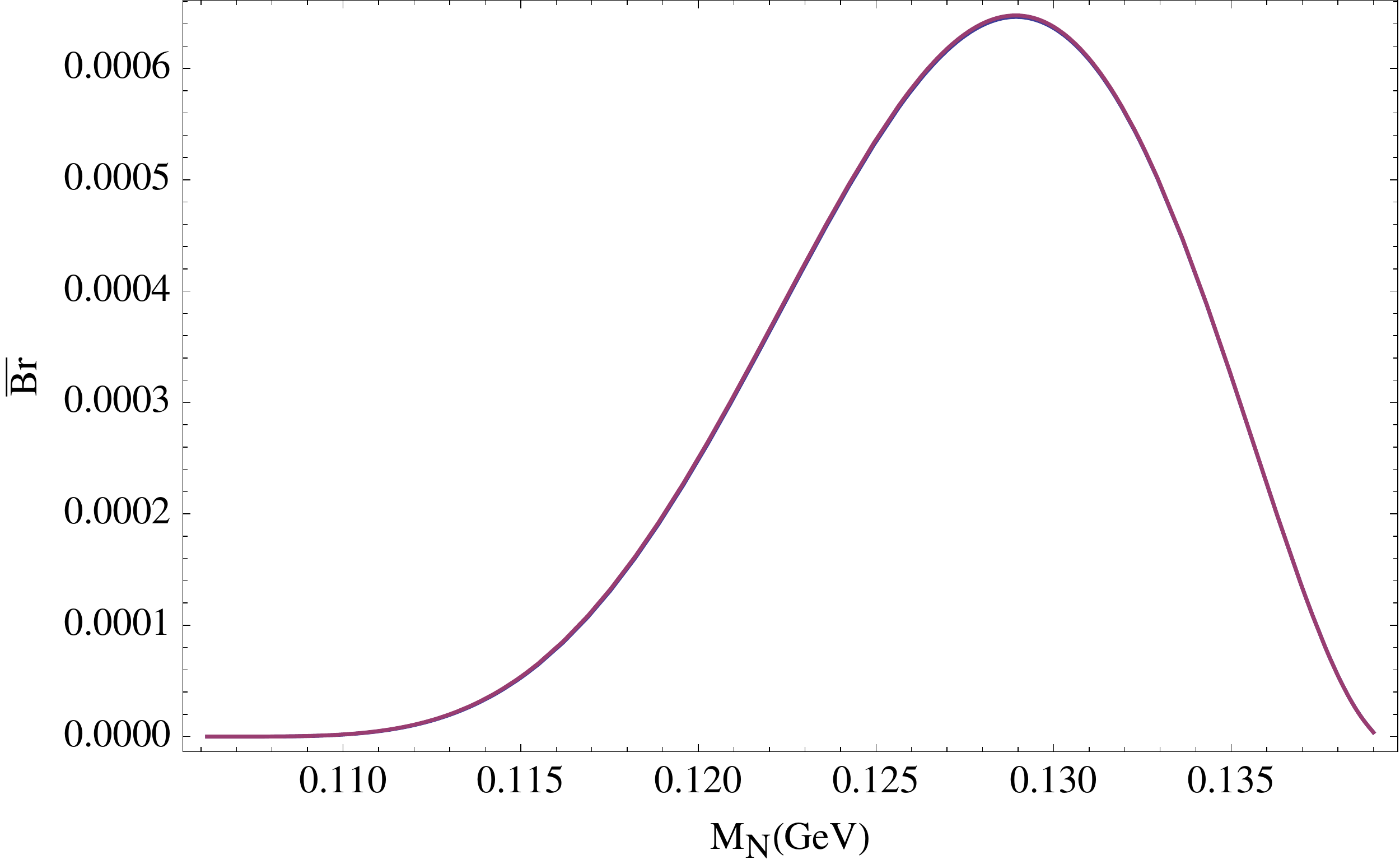}
\vspace{-0.4cm}
\caption{The normalized branching ratio $\Br$, Eq.~(\ref{bBr}),
as a function of the mass
$M_{N_1} \approx M_{N_2} \equiv M_N$. The full formula was used
(with $M_e=0.511 \times 10^{-3}$ GeV). The formula for $M_e=0$ case
gives a line which is in this Figure
indistinguishable from the depicted line.}
\label{bBrfig}
\end{figure}
\begin{figure}[htb] 
\begin{minipage}[b]{.49\linewidth}
\centering\includegraphics[width=73mm]{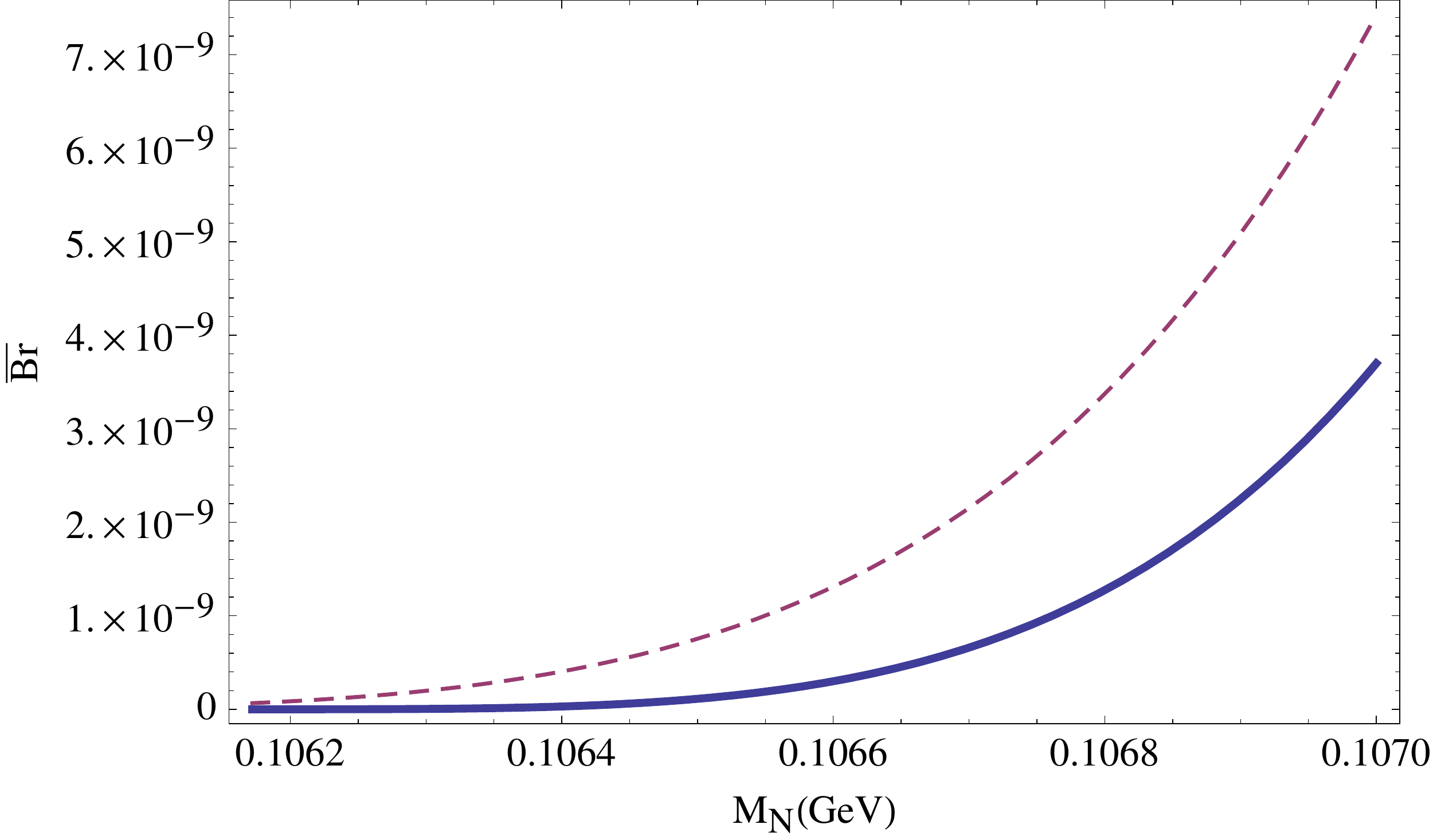}
\end{minipage}
\begin{minipage}[b]{.49\linewidth}
\centering\includegraphics[width=73mm]{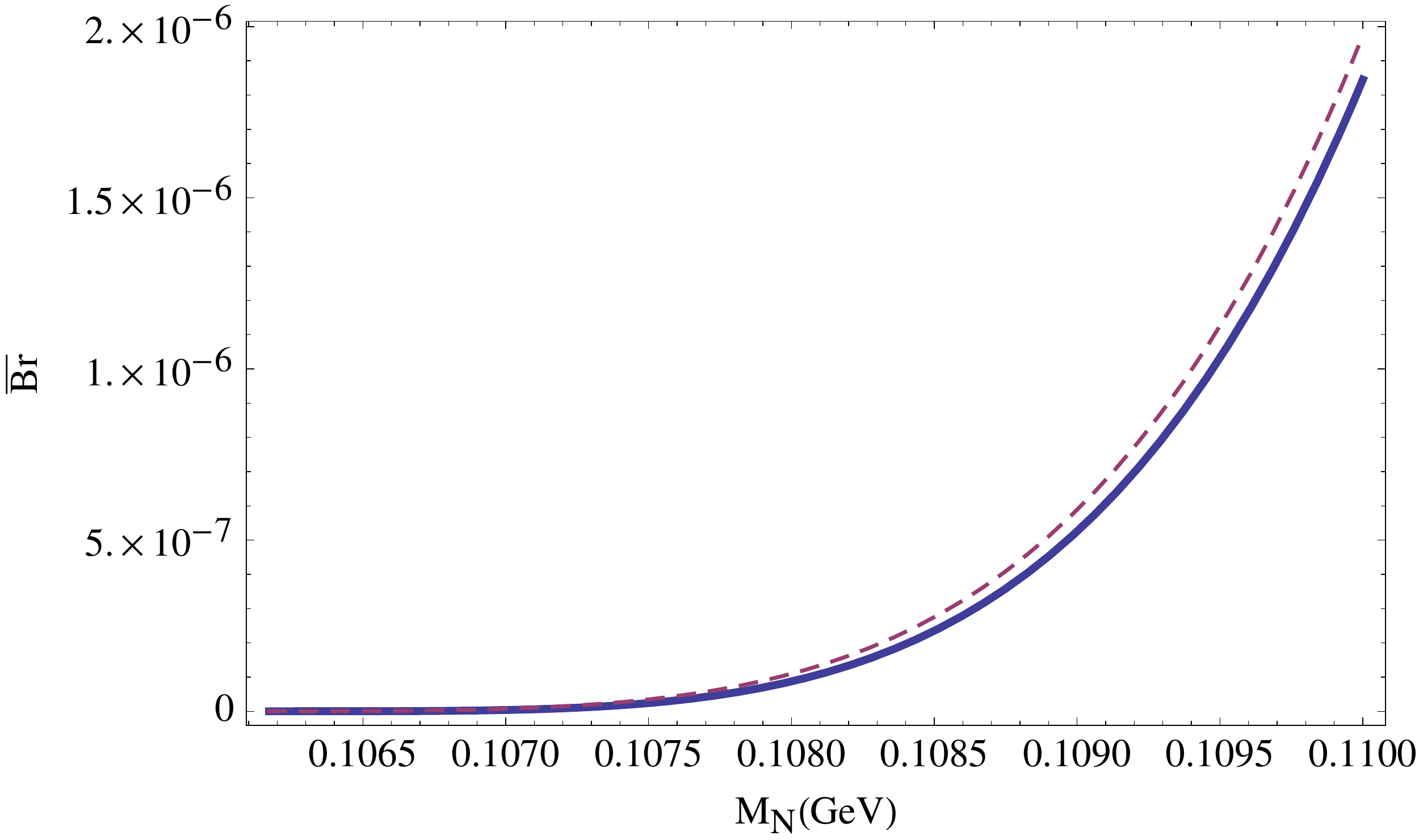}
\end{minipage}
\vspace{-0.4cm}
 \caption{\footnotesize The normalized branching ratio $\Br$ near the
lower end point $M_{\mu} +M_e$ ($=0.1062$ GeV): (a) in the interval below
$0.107$ GeV; (b) in the interval below $0.110$ GeV. The dashed line is
for $M_e=0$, the full line includes
the effects of $M_e = 0.511 \times 10^{-3}$ GeV.}
\label{bBrendfig}
 \end{figure}
We present in Fig.~\ref{bBrfig}
the normalized quantity $\Br$ as a function of $M_N$ in the on-shell kinematic interval
(\ref{MNjint}); and in Figs.~\ref{bBrendfig} the same curve near the lower
end point $M_N \approx M_{\mu}+M_e$ ($=0.1062$ GeV), where the effects of $M_e \not= 0$
are relatively appreciable. Further, in Fig.~\ref{etdelfig} we present
the curves of the overlap suppression factors $\eta(y)/y$ and $\delta(y)$, as a
function of the $N_1$-$N_2$ overlap parameter $y \equiv \Delta M_N/\Gamma_N$.
On the other hand, the (CP asymmetry) branching ratio  
${\rm Br}_{-}$ in the case of mixing one and
maximal CP phases 
(i.e., when $B_{\ell N_j}=1$ for all $\ell$, and $\sin \theta^{(X)}=1$;
${\rm Br}_{-}^{\rm (Di)} = {\rm Br}_{-}^{\rm (Ma)} \equiv {\rm Br}_{-}$ then),
as a function of $\Delta M_N$, is presented in Fig.~\ref{Brfig}.
In that Figure, no overlap effects appear at the values of $\Delta M_N$ presented,
i.e., $\eta=1$.
\begin{figure}[htb]
\centering\includegraphics[width=110mm]{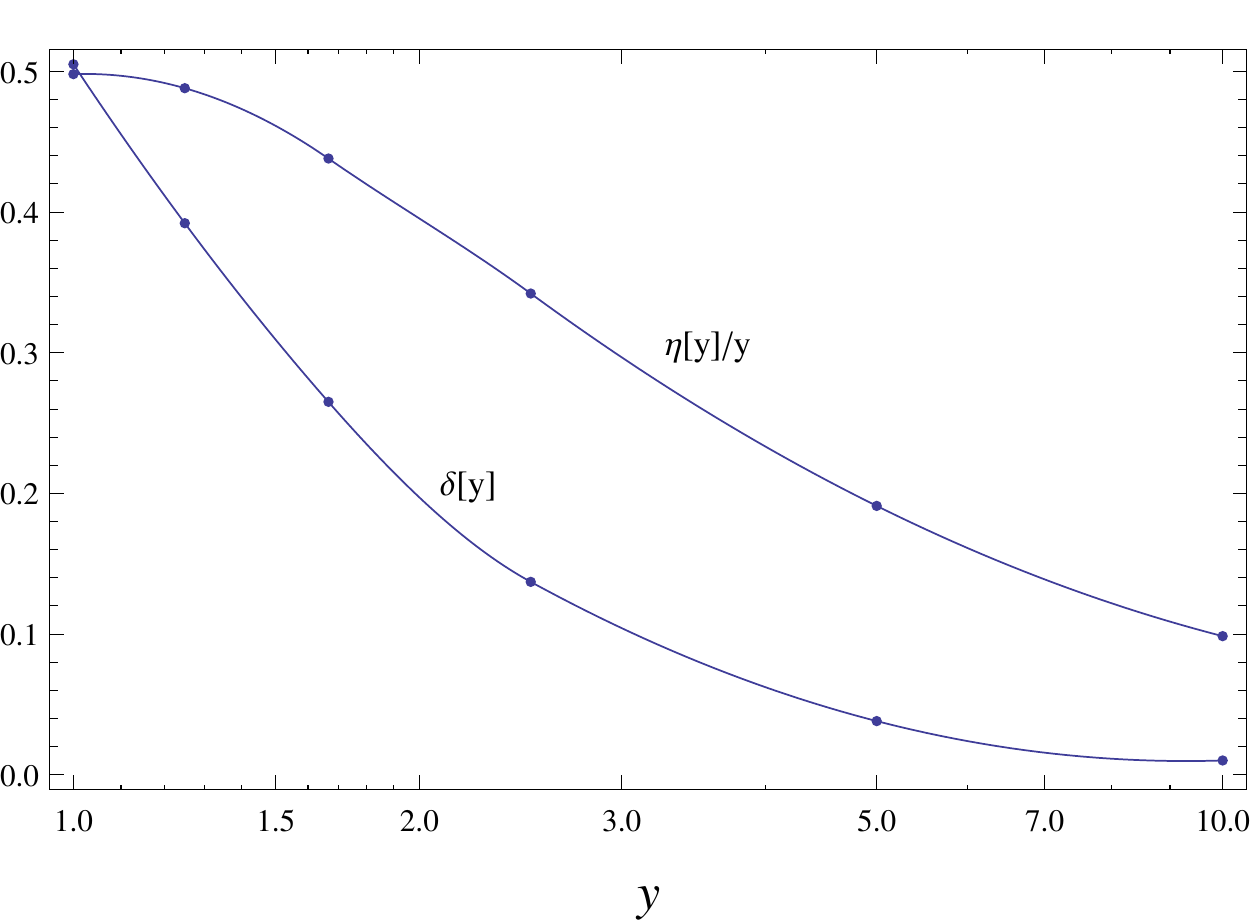}
\vspace{-0.4cm}
\caption{The suppression factors $\eta(y)/y$ and $\delta(y)$, due to the
overlap of the $N_1$ and $N_2$ resonances, as a function of $y \equiv \Delta M_N/\Gamma_N$,
for $1 < y < 10$.}
\label{etdelfig}
\end{figure}

\begin{figure}[htb]
\centering\includegraphics[width=110mm]{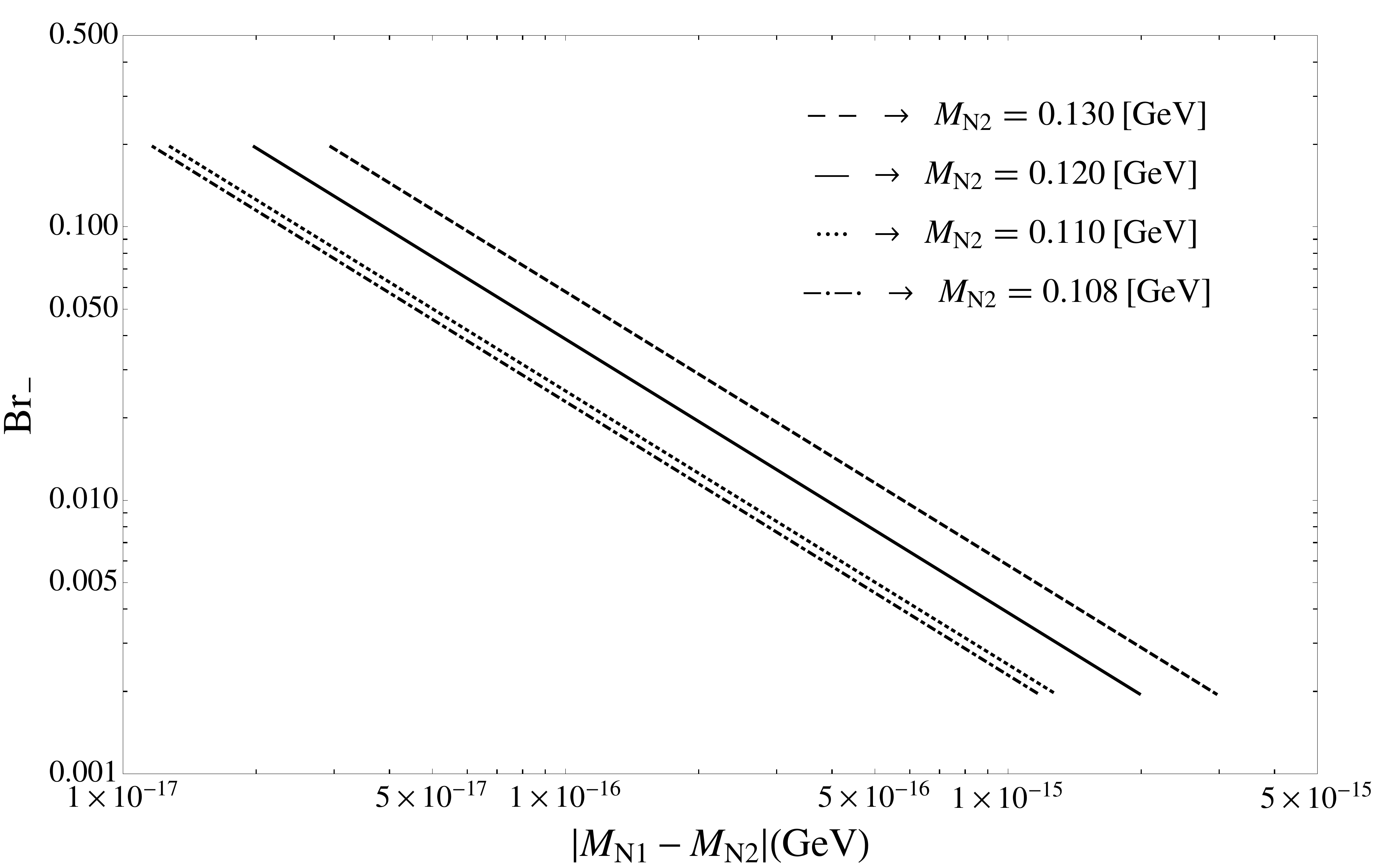}
\vspace{-0.4cm}
\caption{The (CP asymmetry) branching ratio ${\rm Br}_{-}$
as a function of $\Delta M_N = M_{N_2} - M_{N_1}$,
for mixing one ($B_{\ell N_j}=1$) and large CP-violating phases
($\sin \theta^{(X)}=1$), for four different values of $M_{N_2}$.
No suppression effects from the overlap of the
$N_1$ and $N_2$ resonances appear here ($\eta=1$).}
\label{Brfig}
\end{figure}

Therefore, when $y \equiv \Delta M_N/\Gamma_N < 5$, i.e., in
the almost degenerate case of two on-shell neutrinos $N_j$, we can expect
in general the CP asymmetry ratio ${\cal A}_{\rm CP}$ 
of the considered rare process to be $\sim 1$.
The branching ratio for this process, in the case of one $N$ neutrino, 
was considered in Ref.~\cite{JHEP},\footnote{
It was considered in the $M_e=0$ limit, but the general conclusions remain
unchanged with respect to the $M_e = 0.511$ MeV case.}
and all the conclusions about the measurability of this branching ratio
${\rm Br} \approx (1/2) {\rm Br}_{+}$
can be translated into the conclusions about the measurability of the
(CP asymmetry) branching ratio ${\rm Br}_{-}$
in the described almost degenerate scenario,
provided that $|\theta^{\rm (LC)}|, |\theta^{\rm (LV)}| \sim 1$.

This means that the
CP asymmetries could be measured in the future pion factories in
the described scenarios,
provided that the heavy-light mixing parameters $|B_{\ell N_j}|^2$ ($\ell = e, \mu$)
are not many orders of magnitude below the present experimental upper bounds.
The present experimental bounds of the mixing parameters
$|B_{\ell N_j}|^2$ ($\ell = e, \mu, \tau$) in the considered mass range (\ref{MNjint}),
are: $|B_{e N_j}|^2 \stackrel{<}{\sim} 10^{-8}$ \cite{PIENU:2011aa}; $|B_{\mu N_j}|^2 \stackrel{<}{\sim} 10^{-6}$
\cite{BmuN}; $|B_{\tau N_j}|^2 \stackrel{<}{\sim} 10^{-4}$ \cite{BtauN};
cf. also Refs.~\cite{Atre,Ruchayskiy:2011aa}.

The future pion factories, among them the Project X at Fermilab,
will produce charged pions with lab energies $E_{\pi}$ of a few GeV
(i.e., the time dilation factor $\gamma_{\pi} \sim 10^1$),
and luminosities $\sim 10^{22} \ {\rm cm}^{-2} {\rm s}^{-1}$ \cite{ProjX,Geer},
hence, $\sim 10^{29}$ charged pions could be expected per year.
The probability of (on-shell)
neutrino $N$ to decay inside a detector of length $L \sim 10^1$ m in such
pion factories is
\be
P_N \sim \frac{L}{\gamma_{\pi} \tau_N} = \frac{L \Gamma_N}{\gamma_{\pi}} \sim \frac{10^{-2}}{\gamma_{\pi}} \K \sim 10^{-3} \K \ ,
\label{PN}
\ee
where $\K \sim \K_j \propto |B_{\ell N_j}|^2$.
We should multiply the obtained branching ratios ${\rm Br}_{\pm}$
by such acceptance factors $P_N$ to obtain the effective branching
ratios ${\rm Br}_{\pm}^{\rm (eff)}$.

If the largest among the
mixing elements $|B_{\ell N_j}|^2$ ($\ell=e,\mu$)
are $|B_{\mu N_j}|^2$ ($\sim |B_{\mu N}|^2$) ($j=1,2$),
i.e., if we have $|B_{\mu N}|^2 \gg |B_{e N_j}|^2$  ($\sim |B_{e N}|^2$),
the formulas (\ref{PN}) with (\ref{Brpl}) and (\ref{Brmi}) give
\bes
\label{mudom}
\ba
P_N {\rm Br}_{+}^{\rm (Di,Ma)} & \sim & 10^{-3} |B_{e N}|^2 |B_{\mu N}|^2 \Br(M_N) \sim |B_{e N}|^2 |B_{\mu N}|^2 10^{-7} \ ,
\label{mudomBr}
\\
P_N {\rm Br}_{-}^{\rm (Di,Ma)} & \sim &  10^{-3} |B_{e N}|^2 |B_{\mu N}|^2 \sin \theta^{(X)} \Br(M_N)
\sim
|B_{e N}|^2 |B_{\mu N_j}|^2  \sin \theta^{\rm (LC)} 10^{-7} \ .
\label{mudomA}
\ea
\ees
In these relations, we took into account that the LC process dominates
over the LV process in the considered case,
and that $\Br \sim 10^{-4}$ in most of the
on-shell interval for the masses
$M_{N_1} \approx M_{N_2} \equiv M_N$, cf.~Fig.~\ref{bBrfig}.
If in this case, in addition, $|B_{\ell N_j}|^2$ ($\ell=e,\mu$) are
close to their present upper bounds,
$|B_{e N_j}|^2 \sim 10^{-8}$ and $|B_{\mu N_j}|^2 \sim 10^{-6}$,
this implies that $P_N {\rm Br}_{+} \sim 10^{-21}$ and
$P_N {\rm Br}_{-} \sim 10^{-21}$ (the latter provided $\sin \theta^{(X)} \sim 1$),
implying that $\sim 10^8$ events can be detected per year, with the
difference between $\pi^-$ and $\pi^+$ decays also of the order
$\sim 10^8$. This number decreases in proportionality with the factor
$|B_{e N}|^2 |B_{\mu N}|^2$ when this factor decreases.
In this scenario there is almost no difference between the case when
$N_j$ are Dirac and the case when $N_j$ are Majorana.

On the other hand, if the largest among the
mixing elements $|B_{\ell N_j}|^2$ ($\ell=e,\mu$)
are $|B_{e N_j}|^2$ ($\sim |B_{e N}|^2$) ($j=1,2$),
i.e., if we have $|B_{e N}|^2 \gg |B_{\mu N_j}|^2$  ($\sim |B_{\mu N}|^2$),
the formulas (\ref{PN}) with (\ref{Brpl}) and (\ref{Brmi}) give
\bes
\label{edom}
\ba
P_N {\rm Br}_{+}^{\rm (Di)} & \sim & 10^{-3} |B_{e N}|^2 |B_{\mu N}|^2 \Br(M_N) \sim |B_{e N}|^2 |B_{\mu N}|^2 10^{-7} \ ,
\label{mudomBrDi}
\\
P_N {\rm Br}_{+}^{\rm (Ma)} & \sim & 10^{-3} |B_{e N}|^4 \Br(M_N) \sim |B_{e N}|^4 10^{-7} \ ,
\label{mudomBrMa}
\\
P_N {\rm Br}_{-}^{\rm (Di)} & \sim &  10^{-3} |B_{e N}|^2 |B_{\mu N}|^2 \sin \theta^{\rm (LC)} \Br(M_N)
\sim
|B_{e N}|^2 |B_{\mu N}|^2  \sin \theta^{\rm (LC)} 10^{-7} \ .
\label{mudomADi}
\\
P_N {\rm Br}_{-}^{\rm (Ma)} & \sim &  10^{-3} |B_{e N}|^4 \sin \theta^{\rm (LV)} \Br(M_N) \sim
|B_{e N}|^4  \sin \theta^{\rm (LV)} 10^{-7} \ .
\label{mudomAMa}
\ea
\ees
In this considered case, the LV process dominates over the LC process.
If in this case, in addition, $|B_{e N_j}|^2$ are
close to their present upper bounds,
$|B_{e N_j}|^2 \sim 10^{-8}$ (and $|B_{\mu N}|^2 \ll |B_{e N_j}|^2$),
this implies that $P_N {\rm Br}_{+}^{\rm (Ma)} \sim 10^{-23}$ 
($\gg P_N {\rm Br}_{+}^{\rm (Di)}$) and
$P_N {\rm Br}_{-}^{\rm (Ma)} \sim 10^{-23}$ ($\gg P_N {\rm Br}_{-}^{\rm (Di)}$),
assuming that $\sin \theta^{\rm (LV)} \sim 1$.
This implies that $\sim 10^6$ events can be detected per year, with the
difference between $\pi^-$ and $\pi^+$ decays also of the order
$\sim 10^6$, if $N_j$ are Majorana neutrinos (and less events if
$N_j$ are Dirac neutrinos). This number decreases in proportionality with the factor
$|B_{e N}|^4$ when this factor decreases.
In this scenario there is a clear  difference between the case when
$N_j$ are Dirac and the case when $N_j$ are Majorana.

The mentioned present experimental upper bounds on the mixings
($|B_{e N_j}|^2 \stackrel{<}{\sim} 10^{-8}$; $|B_{\mu N_j}|^2 \stackrel{<}{\sim} 10^{-6}$)
suggest that the first of the mentioned two scenarios is more probable,
i.e., that the LC processes dominate over the LV processes.

The measurement of the CP asymmetries alone cannot
distinguish between the Dirac and the Majorana character of intermediate
neutrinos $N_j$'s. However, as argued in Ref.~\cite{JHEP}, the neutrino
character could be determined from the measured differential decay rates
of these processes with respect to the muon energy
$E_{\mu}$ in the $N_j$ rest frame, $d \Gamma/d E_{\mu}$,
if the heavy-light mixing elements satisfy the relation $|B_{e N_j}|
\stackrel{>}{\sim} |B_{\mu N_j}|$ (if $|B_{e N_j}| \ll |B_{\mu N_j}|$, the $LC$ process dominates).

\section{Summary}
\label{sec:concl}

We investigated the rare decays of charged pions,
$\pi^{\pm} \to e^\pm N_j \to e^{\pm} e^{\pm} \mu^{\mp} \nu$, in scenarios with two heavy
sterile neutrinos $N_j$ ($j=1,2$).
Such scenarios allow the mentioned decays to proceed with exchange of on-shell
intermediate neutrinos at the tree level, but are suppressed by
the heavy-light neutrino mixing elements of the PMNS matrix. The mentioned decays can be
of the lepton-number-conserving (LC) type ($\nu = \nu_{e}, {\bar \nu}_e$),
or of the lepton-number-violating (LV) type ($\nu={\bar \nu}_{\mu}, \nu_{\mu}$).
If the $N_j$ neutrinos are of Dirac nature, only LC decays take place; if
they are of Majorana nature, both LC and LV decays take place.
In Ref.~\cite{JHEP} such processes were studied with a view to
ascertain the nature of the intermediate neutrino $N_j$, and it was shown
there that it may be possible to do this in the future pion factories where
the number of produced charged pions will be exceedingly high. In the
present work, on the other hand, we investigated the possibility to
ascertain the CP violation in such processes. Such a CP violation
originates from the interference between the $N_1$ and $N_2$ exchange processes
and the existence of possible CP-violating phases in the PMNS
mixing matrix.
We showed that such signals of CP violation could be detected in the future
pion factories if we have (at least) two sterile neutrinos in the mentioned
mass interval and such that their masses are almost degenerate,
i.e., when the mass difference $\Delta M_N$ between them is not many orders
of magnitude larger than their decay width $\Gamma_N$. 
Therefore, our calculation suggests that the observation of CP violation
in pion decays would be consistent with the existence of $\nu$MSM model
\cite{nuMSM,Shapo,nuMSMrev}, with the two almost degenerate heavy neutrinos
in the lower mass range of the model.
The Majorana nature of the neutrinos offers more possibilities of
CP violation because there are more  CP-violating phases in the PMNS matrix
than in the case when the neutrinos have Dirac nature. On the other hand,
the present experimental bounds on the heavy-light mixings allow higher
rates and more appreciable CP-violating effects in these processes
in the LC channels than in the LV channels, i.e., in the scenarios where
the Majorana nature of the neutrinos is difficult to discern.

\acknowledgments{
This work was supported in part by (Chile) FONDECYT Grant No.~1130599 (G.C. and C.S.K.), by CONICYT Fellowship ``Beca de Doctorado Nacional'' and Proyecto PIIC 2013 (J.Z.S.).
The work of C.S.K. was supported by the NRF
grant funded by the Korean Government of the MEST
(No. 2011-0017430) and (No. 2011-0020333).}

\appendix
\section{Explicit formulas}
\label{app1}

The squared matrix element $| {\cal T}^{(X)}(\pi^{\pm}) | ^2$ in Eq.~(\ref{GX1}),
where X=LV, LC,
is a combination of contributions from $N_1$ and $N_2$ and
from the two channels $D$ (direct) and $C$ (crossed)
\ba
\lefteqn{
| {\cal T}^{(X)}(\pi^{\pm}) | ^2 =
K^2 \sum_{i=1}^2 \sum_{j=1}^2 k_{i,\pm}^{(X) *} k_{j,\pm}^{(X)} }
\nonumber\\
&&
\times {\bigg [} P_i^{(X)}(D) P_j^{(X)}(D)^{*} T_{\pm}^{(X)}(DD^{*})
+  P_i^{(X)}(C) P_j^{(X)}(C)^{*} T_{\pm}^{(X)}(CC^{*})
\nonumber\\
&&+
\left( P_i^{(X)}(D) P_j^{(X)}(C)^{*} T_{\pm}^{(X)}(DC^{*}) +
P_i^{(X)}(C) P_j^{(X)}(D)^{*} T_{\pm}^{(X)}(CD^{*}) \right) {\bigg ]},
\label{calTX}
\ea
where the constant $K^2$ is given in Eq.~(\ref{Ksqr}),
the mixing  factors $k_{j,\pm}^{(X)}$ in Eq.~(\ref{kj}),
and $P_j^{(X)}(Y)$  represent the  $N_j$ propagator functions
Eq.~(\ref{Pj}) of the direct and crossed channels ($Y=D, C$).

Here we write down the explicit formulas for the direct ($DD^{*}$), crossed ($CC^{*}$)
and direct-crossed interference
($DC^{*}$, $CD^{*}$) elements 
[$T_{\pm}^{(X)}( DD^{*})$, $T_{\pm}^{(X)}( CC^{*})$,  $T_{\pm}^{(X)}( DC^{*})$,
$T_{\pm}^{(X)}( CD^{*})$] appearing in Eqs.~(\ref{calTX}). For the lepton number violating
(LV) process of Fig.~\ref{FigLV}, these are
\bes
\label{TLV}
\ba
T_{\pm}^{\rm (LV)}(DD^{*})&=& 256 (p_2 \cdot p_{\nu}) \left[
- M_{\pi}^2 (p_1 \cdot p_{\mu}) + 2 (p_1 \cdot p_{\pi}) (p_{\mu} \cdot p_{\pi}) \right] \equiv T^{\rm (LV)}(DD^{*})  \ ,
\label{TLVDDp}
\\
T_{\pm}^{\rm (LV)}(CC^{*})&=& 256 (p_1 \cdot p_{\nu}) \left[
- M_{\pi}^2 (p_2 \cdot p_{\mu}) + 2 (p_2 \cdot p_{\pi}) (p_{\mu} \cdot p_{\pi}) \right] \equiv T^{\rm (LV)}(CC^{*}) \ ,
\label{TLVCCp}
\\
T_{\pm}^{\rm (LV)}(DC^{*})&=& 128 {\bigg \{}
(p_1 \cdot p_{\nu})
\left[ M_{\pi}^2 (p_2 \cdot p_{\mu}) - 2 (p_2 \cdot p_{\pi})(p_{\mu} \cdot p_{\pi}) \right]
\nonumber\\
&&+
(p_2 \cdot p_{\nu})
\left[ M_{\pi}^2 (p_1 \cdot p_{\mu}) - 2 (p_1 \cdot p_{\pi})(p_{\mu} \cdot p_{\pi}) \right]
\nonumber\\
&&- (p_1 \cdot p_2)
\left[ M_{\pi}^2 (p_{\nu} \cdot p_{\mu}) - 2 (p_{\nu} \cdot p_{\pi})(p_{\mu} \cdot p_{\pi}) \right] {\bigg \}}
\nonumber\\
&&
\mp i {\bigg \{}
- (p_1 \cdot p_{\pi}) \epsilon(p_2,p_{\nu},p_{\mu},p_{\pi})
+ (p_2 \cdot p_{\pi}) \epsilon(p_1,p_{\nu},p_{\mu},p_{\pi})
\nonumber\\
&&- (p_{\nu} \cdot p_{\pi}) \epsilon(p_1,p_2,p_{\mu},p_{\pi})
- (p_{\mu} \cdot p_{\pi}) \epsilon(p_1,p_2,p_{\nu},p_{\pi}) {\bigg \}} \ ,
\label{TLVDCp}
\\
T_{\pm}^{\rm (LV)}(CD^{*})&=& \left( T_{\pm}^{\rm (LV)}(DC^{*}) \right)^{*}
= T_{\mp}^{\rm (LV)}(DC^{*}) \ ,
\label{TLVCDp}
\ea
\ees
where we denoted
\be
\epsilon(q_1, q_2, q_3, q_4) \equiv \epsilon^{\eta_1 \eta_2 \eta_3 \eta_4}
(q_1)_{\eta_1} (q_2)_{\eta_2} (q_3)_{\eta_3} (q_4)_{\eta_4} \ ,
\label{eps}
\ee
and $\epsilon^{\eta_1 \eta_2 \eta_3 \eta_4}$ is the totally antisymmetric Levi-Civita tensor with the
sign convention $\epsilon^{0123}=+1$.

For the lepton number conserving (LC) process of Fig.~\ref{FigLC},
the corresponding expressions are
\bes
\label{TLC}
\ba
T_{\pm}^{\rm (LC)}(DD^{*})&=& 256 (p_{\mu} \cdot p_{\nu}) {\bigg [}
(p_1 \cdot p_2) \left( M_{\pi}^4 - M_{\pi}^2 M_e^2
- 4 M_{\pi}^2 (p_1 \cdot p_{\pi}) + 4 (p_1 \cdot p_{\pi})^2 \right)
\nonumber\\ &&
+ 2 M_e^2 (p_2 \cdot p_{\pi}) (M_{\pi}^2 - p_1 \cdot p_{\pi}) {\bigg ]} 
\equiv T^{\rm (LC)}(DD^{*}) \ ,
\label{TLCDDp}
\\
T_{\pm}^{\rm (LC)}(CC^{*})&=& 256 (p_{\mu} \cdot p_{\nu}) {\bigg [}
(p_1 \cdot p_2) \left( M_{\pi}^4 - M_{\pi}^2 M_e^2
- 4 M_{\pi}^2 (p_2 \cdot p_{\pi}) + 4 (p_2 \cdot p_{\pi})^2 \right)
\nonumber\\ &&
+ 2 M_e^2 (p_1 \cdot p_{\pi}) (M_{\pi}^2 - p_2 \cdot p_{\pi}) {\bigg ]}  
\equiv T^{\rm (LC)}(CC^{*}) \ ,
\label{TLCCCp}
\\
T_{\pm}^{\rm (LC)}(DC^{*})&=& 256 (p_{\mu} \cdot p_{\nu})
{\bigg [} (p_1 \cdot p_2) (M_{\pi}^2 - 2 p_1 \cdot p_{\pi}) (M_{\pi}^2 - 2 p_2 \cdot p_{\pi})
\nonumber\\
&& + M_e^2 \left(-2 (p_1 \cdot p_{\pi})^2 -2 (p_2 \cdot p_{\pi})^2
+ M_{\pi}^2 (p_1 +p_2) \cdot p_{\pi} + M_{\pi}^2 M_e^2 \right) {\bigg ]}
\equiv T^{\rm (LC)}(DC^{*}) \ ,
\label{TLCDCp}
\\
T^{\rm (LC)}(CD^{*})&=& \left( T^{\rm (LC)}(DC^{*}) \right)^{*} \ .
\label{TLCCDp}
\ea
\ees

On the basis of these expressions, and the corresponding definitions
of the normalized (i.e., without mixings) decay width matrices
$\bG_{\pm}^{(X)}(YZ^{*})$ ($X=$LV, LC; $Y, Z = D, C$) in Eq.~(\ref{GXij}),
and using the symmetry of the $d_4$ integration under the
exchange $p_1 \leftrightarrow p_2$ (this because: $M_1=M_2=M_e$ in our
considered case),
the following symmetry relations hold between the
elements of the  decay width matrices of Eq.~(\ref{GXij}):
\bes
\label{symm}
\ba
\bG^{(X)}(DD^{*})_{ij} & = & \bG^{(X)}(CC^{*})_{ij} \ , \quad
\bG^{(X)}(DD^{*})_{ji} = \left( \bG^{(X)}(DD^{*})_{ij} \right)^{*} \ ,
\label{symmGDD}
\\
\bG_{\pm}^{(X)}(CD^{*})_{ij} & = & \bG_{\pm}^{(X)}(DC^{*})_{ij} =
 \left( \bG_{\pm}^{(X)}(CD^{*})_{ji} \right)^{*} \ .
\label{symmCD}
\ea
\ees

\section{Explicit formula for $\bG^{(X)}$ when $M_e \not= 0$}
\label{app2}

The formula (\ref{GXDDp}) is obtained by performing the integration
of the differential decay width $d \bG^{\rm (LV)}/d E_{\mu}$ over the muon
energy $E_{\mu}$, in the rest frame of the $N_j$ neutrino.
The expression for $d \bG^{\rm (LV)}/d E_{\mu}$ is
written explicitly, e.g., in Appendix A of Ref.~\cite{JHEP}.
This gives
\bes
\label{intDGDemu}
\ba
\lefteqn{
\bG^{\rm (LV)}(DD^{*})_{jj} =  K^2 \; \frac{1}{2!} \frac{1}{2 M_{\pi} (2 \pi)^8}
\int d_4 \; |P_j^{\rm (LV)}(D)|^2 \; T^{\rm (LV)}(DD^{*})
}
\label{Gjj1}
\\
& = &  K^2 \; \frac{1}{2!} \frac{1}{2 M_{\pi} (2 \pi)^8} \int
d_4 \; \frac{\pi}{M_{N_j} \Gamma_{N_j}}
\delta \left( (p_{\pi} \! - \! p_1)^2 - M_{N_j}^2
\right) M_{N_j}^2 \; T^{\rm (LV)}(DD^{*}) = \cdots
\label{Gjj2}
\\
& = &  K^2 \; \frac{1}{(2 \pi)^4} \frac{M_{N_j}}{\Gamma_{N_j} M_{\pi}^3}
\lambda^{1/2}(M_{\pi}^2,M_{N_j}^2,M_e^2) \times
\frac{1}{2 M_{N_j}}
\left[ M_{\pi}^2 (M_{N_j}^2 + M_e^2) - (M_{N_j}^2 - M_e^2)^2 \right]
\nonumber\\
&& \times
\int_{M_{\mu}}^{(M_{N_j}^2 + M_{\mu}^2 - M_e^2)/(2 M_e)}
d E_{\mu} E_{\mu} \sqrt{E_{\mu}^2 - M_{\mu}^2}
\frac{ ( M_{N_j}^2 - 2 M_{N_j} E_{\mu} + M_{\mu}^2 - M_e^2)^2}
{(M_{N_j}^2 - 2 M_{N_j} E_{\mu} + M_{\mu}^2)} \ .
\label{Gjj3}
\ea
\ees
The ellipses in Eq.~(\ref{Gjj2}) indicate the analytic
integrations over the four-particle final phase space of the process
of Fig.~\ref{FigLV}(a) with the exception of $E_{\mu}$
(in the rest frame of $N_j$), performed in Ref.~\cite{JHEP}.
Eq.~(\ref{Gjj3}) then uses the differential decay width $d \bG^{\rm (LV)}(DD^{*})/d E_{\mu}$
obtained in Ref.~\cite{JHEP}.\footnote{
Eq.~(A.7) of that reference, with the corresponding
replacements: $m_M \mapsto M_{\pi}$, $m_N \mapsto M_{N_j}$, $m_{\ell} \mapsto M_{\mu}$,
$m_1=m_2 \mapsto M_e$.} The integration in Eq.~(\ref{Gjj3}) can be performed explicitly
(in Ref.~\cite{JHEP} it was performed only in the limit $M_e=0$), and the result
is Eq.~(\ref{GXDDp}) with notations (\ref{notGXDDp}) and the
function ${\cal F}(x_j,x_{ej})$ given explicitly here
\ba
\lefteqn{
{\cal F}(x_j,x_{ej}) =
{\Bigg \{}
\lambda^{1/2} (1, x_j, x_{ej}) {\big [} (1 + x_j) (1 -8 x_j + x_j^2)  -
x_{ej} (7 - 12 x_j + 7 x_j^2)
}
\nonumber\\
&&
- 7 x_{ej}^2 (1 + x_j)  + x_{ej}^3 {\big ]}
- 24 (1 - x_{ej}^2) x_j^2 \ln 2
\nonumber\\
&&
+  12 {\bigg [} - x_j^2 (1 - x_{ej}^2) \ln x_j
+ (2 x_j^2 -x_{ej}^2 (1 + x_j^2)) \ln (1 + x_j
+ \lambda^{1/2} (1, x_j, x_{ej})  - x_{ej})
\nonumber\\
&&
+ x_{ej}^2 (1 - x_j^2)
\ln \left( \frac{(1 - x_j)^2 + (1-x_j) \lambda^{1/2} (1, x_j, x_{ej}) - x_{ej} (1+x_j)}{x_{ej}}
\right) {\bigg ]}
{\Bigg \}} \ ,
\label{calF}
\ea
It turns out that the integration over the differential decay width of the
lepton number conserving case, $d \bG^{\rm (LC)}/d E_{\mu}$,
\ba
\lefteqn{
\bG^{\rm (LC)}(DD^{*})_{jj} =
K^2 \; \frac{1}{(2 \pi)^4} \frac{M_{N_j}}{\Gamma_{N_j} M_{\pi}^3}
\lambda^{1/2}(M_{\pi}^2,M_{N_j}^2,M_e^2)
\times \frac{1}{96 M_{N_j}^2}
}
\nonumber\\
&& \times
\int_{M_{\mu}}^{(M_{N_j}^2 + M_{\mu}^2 - M_e^2)/(2 M_e)} d E_{\mu}
\frac{1}{ \left[ M_{\mu}^2 + M_{N_j} (-2 E_{\mu} + M_{N_j}) \right]^3}
\nonumber\\
&& \times {\bigg \{} 8 \sqrt{(E_{\mu}^2 - M_{\mu}^2)} M_{N_j}
\left[ (2 E_{\mu} - M_{N_j}) M_{N_j} - M_{\mu}^2 + M_e^2 \right]^2
\left[M_{\pi}^2 M_{N_j}^2  -  M_{N_j}^4 + M_e^2 (M_{\pi}^2+2 M_{N_j}^2)  -  M_e^4
\right]
\nonumber\\
&& \
\times
{\Big [}  8 E_{\mu}^3 M_{N_j}^2 - 2 M_{\mu}^2 M_{N_j} (M_{\mu}^2 +M_{N_j}^2+ 2 M_e^2)
 - 2 E_{\mu}^2 M_{N_j} \left( 5 (M_{\mu}^2 + M_{N_j}^2)+M_e^2 \right)
\nonumber\\
&&
+ E_{\mu}
\left( 3 M_{\mu}^4 + 10 M_{\mu}^2 M_{N_j}^2 + 3 M_{N_j}^4 + 3 M_e^2 (M_{\mu}^2 + M_{N_j}^2) \right) {\Big ]}
{\bigg \}}
\label{GLCjj}
\ea
gives the same result as the $X=LV$ case, i.e., Eqs.~(\ref{GXDDp}) with
(\ref{calF}).
In Eq.~(\ref{GLCjj}) we inserted the differential
decay width $d \bG^{\rm (LC)}(DD^{*})_{jj}/d E_{\mu}$ as obtained
in Eq.~(A.14) of Ref.~\cite{JHEP}.\footnote{
There is a typo in Eq.~(A.16) of Ref.~\cite{JHEP}, i.e., in the
expression for $d \bG^{\rm (LC)}(DD^{*})_{jj}/d E_{\mu}$ in the $M_e=0$ limit:
in the second line of that equation,
$E_{\ell}^2$ must be replaced by $2 E_{\ell}^2$.}


\begin{thebibliography}{99}

\bibitem{0nubb}
  G.~Racah,
  Nuovo Cim.\  {\bf 14}, 322 (1937);
  W.~H.~Furry,
  Phys.\ Rev.\  {\bf 56}, 1184 (1939);
H.~Primakoff and S.~P.~Rosen, Rep. Prog. Phys. {\bf 22}, 121 (1959);
\emph{ibid.}
  Phys.\ Rev.\  {\bf 184}, 1925 (1969);
\emph{ibid.}
  Ann.\ Rev.\ Nucl.\ Part.\ Sci.\  {\bf 31}, 145 (1981);
  J.~Schechter and J.~W.~F.~Valle,
  Phys.\ Rev.\ D {\bf 25}, 2951 (1982);
 M.~Doi, T.~Kotani and E.~Takasugi,
  Prog.\ Theor.\ Phys.\ Suppl.\  {\bf 83}, 1 (1985);
  S.~R.~Elliott and J.~Engel,
  J.\ Phys.\ G G {\bf 30}, R183 (2004)
  [hep-ph/0405078];
  V.~A.~Rodin, A.~Faessler, F.~Simkovic and P.~Vogel,
  Nucl.\ Phys.\ A {\bf 766}, 107 (2006)
  [Erratum-ibid.\ A {\bf 793}, 213 (2007)]
  [arXiv:0706.4304 [nucl-th]].


\bibitem{scatt}
  W.~-Y.~Keung and G.~Senjanovi\'c,
  Phys.\ Rev.\ Lett.\  {\bf 50}, 1427 (1983);
  V.~Tello, M.~Nemev\v{s}ek, F.~Nesti, G.~Senjanovi\'c and F.~Vissani,
  Phys.\ Rev.\ Lett.\  {\bf 106}, 151801 (2011)
  [arXiv:1011.3522 [hep-ph]];
  M.~Nemev\v{s}ek, F.~Nesti, G.~Senjanovi\'c and V.~Tello,
  arXiv:1112.3061 [hep-ph];
  G.~Senjanovi\'c,
  Riv.\ Nuovo Cim.\  {\bf 034}, 1 (2011).


\bibitem{rmeson}
 L.~S.~Littenberg and R.~E.~Shrock,
  Phys.\ Rev.\ Lett.\  {\bf 68}, 443 (1992);
 {\it ibid.},
  Phys.\ Lett.\ B {\bf 491}, 285 (2000)
  [hep-ph/0005285];
  C.~Dib, V.~Gribanov, S.~Kovalenko and I.~Schmidt,
  Phys.\ Lett.\ B {\bf 493}, 82 (2000)
  [hep-ph/0006277];
  A.~Ali, A.~V.~Borisov and N.~B.~Zamorin,
  Eur.\ Phys.\ J.\ C {\bf 21}, 123 (2001)
  [hep-ph/0104123];
  M.~A.~Ivanov and S.~G.~Kovalenko,
  Phys.\ Rev.\ D {\bf 71}, 053004 (2005)
  [hep-ph/0412198].
  A.~de Gouvea and J.~Jenkins,
  Phys.\ Rev.\ D {\bf 77}, 013008 (2008)
  [arXiv:0708.1344 [hep-ph]];
  G.~Cveti\v{c}, C.~Dib, S.~K.~Kang and C.~S.~Kim,
  Phys.\ Rev.\ D {\bf 82}, 053010 (2010)
  [arXiv:1005.4282 [hep-ph]];
  J.~C.~Helo, S.~Kovalenko and I.~Schmidt,
  Nucl.\ Phys.\ B {\bf 853}, 80 (2011)
  [arXiv:1005.1607 [hep-ph]].

\bibitem{JHEP}
  G.~Cveti\v c, C.~Dib and C.~S.~Kim,
  JHEP {\bf 1206}, 149 (2012)
  [arXiv:1203.0573 [hep-ph]].

\bibitem{Pontecorvo}
  B.~Pontecorvo,
  Sov.\ Phys.\ JETP {\bf 7}, 172 (1958)
  [Zh.\ Eksp.\ Teor.\ Fiz.\  {\bf 34}, 247 (1957)].


\bibitem{oscatm}
  Y.~Fukuda {\it et al.}  [Super-Kamiokande Collaboration],
  Phys.\ Rev.\ Lett.\  {\bf 81}, 1562 (1998)
  [hep-ex/9807003].

\bibitem{oscsol}
  Q.~R.~Ahmad {\it et al.}  [SNO Collaboration],
  Phys.\ Rev.\ Lett.\  {\bf 89}, 011301 (2002)
  [nucl-ex/0204008];
  P.~Lipari,
  Phys.\ Rev.\ D {\bf 64}, 033002 (2001)
  [hep-ph/0102046];
  Z.~Rahman, A.~Dasgupta and R.~Adhikari,
  arXiv:1210.2603 [hep-ph].
  arXiv:1210.4801 [hep-ph].

\bibitem{oscnuc}
  K.~Eguchi {\it et al.}  [KamLAND Collaboration],
  Phys.\ Rev.\ Lett.\  {\bf 90}, 021802 (2003)
  [hep-ex/0212021].


\bibitem{seesaw}
  P.~Minkowski,
  Phys.\ Lett.\ B {\bf 67}, 421 (1977);
M.~Gell-Mann, P.~Ramond and R.~Slansky, in Sanibel Conference, 
``The Family Group in Grand Unified Theories,'' Febr.~1979, CALT-68-700, 
reprinted in hep-ph/9809459;
"Complex Spinors and Unified Theories," Print 80-0576, 
published in: D.~Freedman et al. (Eds.), ``Supergravity'', North-Holland, Amsterdam, 1979;
  T.~Yanagida,
  Conf.\ Proc.\ C {\bf 7902131}, 95 (1979);
S.~L.~Glashow, in: M.~Levy et al. (Eds.), ``Quarks and Leptons,'' Cargese,
Plenum, New York, 1980, p.~707;
  R.~N.~Mohapatra and G.~Senjanovi\'c,
  Phys.\ Rev.\ Lett.\  {\bf 44}, 912 (1980).


\bibitem{Bilenky}
S.~Bilenky, {\it Introduction to the Physics of Massive and Mixed Neutrinos\/},
Lecture Notes in Physics 817, Springer Verlag, Berlin, Heidelberg, 2010.


\bibitem{oscCP}
  N.~Cabibbo,
  Phys.\ Lett.\ B {\bf 72}, 333 (1978);

\bibitem{nuMSM}
  T.~Asaka, S.~Blanchet and M.~Shaposhnikov,
  Phys.\ Lett.\ B {\bf 631}, 151 (2005)
  [hep-ph/0503065];
  T.~Asaka and M.~Shaposhnikov,
  Phys.\ Lett.\ B {\bf 620}, 17 (2005)
  [hep-ph/0505013].

\bibitem{Shapo}
  D.~Gorbunov and M.~Shaposhnikov,
  JHEP {\bf 0710}, 015 (2007)
  [arXiv:0705.1729 [hep-ph]].

\bibitem{PDG2012}
  J.~Beringer {\it et al.}  [Particle Data Group Collaboration],
  Phys.\ Rev.\ D {\bf 86}, 010001 (2012).


 \bibitem{Atre}
  A.~Atre, T.~Han, S.~Pascoli and B.~Zhang,
  JHEP {\bf 0905}, 030 (2009)
  [arXiv:0901.3589 [hep-ph]],
and references therein.

\bibitem{nuMSMrev}
  A.~Boyarsky, O.~Ruchayskiy and M.~Shaposhnikov,
  Ann.\ Rev.\ Nucl.\ Part.\ Sci.\  {\bf 59}, 191 (2009)
  [arXiv:0901.0011 [hep-ph]].

\bibitem{CDSh}
  L.~Canetti, M.~Drewes and M.~Shaposhnikov,
  Phys.\ Rev.\ Lett.\  {\bf 110}, no. 6, 061801 (2013)
  [arXiv:1204.3902 [hep-ph]];
  L.~Canetti, M.~Drewes, T.~Frossard and M.~Shaposhnikov,
  Phys.\ Rev.\ D {\bf 87}, 093006 (2013)
  [arXiv:1208.4607 [hep-ph]].

\bibitem{DM}
  E.~Bulbul, M.~Markevitch, A.~Foster, R.~K.~Smith, M.~Loewenstein and S.~W.~Randall,
  arXiv:1402.2301 [astro-ph.CO];
  A.~Boyarsky, O.~Ruchayskiy, D.~Iakubovskyi and J.~Franse,
  arXiv:1402.4119 [astro-ph.CO].

\bibitem{CERN-SPS}
W. Bonibento {\it et al.}, CERN-SPSC-2013-024, CERN-EOI-010,
[arXiv:1310.1762 [hep-ex]];
R.~Jacobson, ``Search for heavy neutral neutrinos at the SPS,''
presented at {\it High Energy Physics in the LHC Era,\/}, UTFSM,
Valpara\'{\i}so, Chile, December 16-20, 2014,
https://indico.cern.ch/contributionDisplay.py?contribId=215\&confId=252857

\bibitem{PIENU:2011aa}
  M.~Aoki {\it et al.}  [PIENU Collaboration],
  Phys.\ Rev.\ D {\bf 84}, 052002 (2011)
  [arXiv:1106.4055 [hep-ex]].

\bibitem{BmuN}
  T.~Yamazaki, T.~Ishikawa, Y.~Akiba, M.~Iwasaki, K.~H.~Tanaka, S.~Ohtake, H.~Tamura and M.~Nakajima {\it et al.},
  Conf.\ Proc.\ C {\bf 840719}, 262 (1984);
  R.~S.~Hayano, T.~Taniguchi, T.~Yamanaka, T.~Tanimori, R.~Enomoto, A.~Ishibashi, T.~Ishikawa and S.~Sato {\it et al.},
  Phys.\ Rev.\ Lett.\  {\bf 49}, 1305 (1982);
  A.~Kusenko, S.~Pascoli and D.~Semikoz,
  JHEP {\bf 0511}, 028 (2005)
  [hep-ph/0405198].

\bibitem{BtauN}
  J.~Orloff, A.~N.~Rozanov and C.~Santoni,
  Phys.\ Lett.\ B {\bf 550}, 8 (2002)
  [hep-ph/0208075].

\bibitem{Ruchayskiy:2011aa}
  O.~Ruchayskiy and A.~Ivashko,
  JHEP {\bf 1206}, 100 (2012)
  [arXiv:1112.3319 [hep-ph]].

\bibitem{ProjX}
{\it Project X and the Science of
the Intensity Frontier}, white paper based on the
Project X Physics Workshop, Fermilab, USA, 9-10 November 2009
(http://projectx.fnal.gov/pdfs/ProjectXwhitepaperJan.v2.pdf).


\bibitem{Geer}
S.~Geer, private communication, 2012.


\end{thebibliography}
\end{document}